\documentclass[10pt,journal,compsoc]{IEEEtran}

\usepackage{cite}
\usepackage{amsmath,amssymb,amsfonts}
\usepackage{algorithmic}
\usepackage{graphicx}
\usepackage{textcomp}
\usepackage{xcolor}
\usepackage[hyphens]{url}
\usepackage{booktabs}
\usepackage{multirow}
\usepackage{array}
\usepackage{pifont}
\newcommand{\cmark}{\ding{51}}
\newcommand{\xmark}{\ding{55}}

\usepackage{tikz}
\usepackage{pgfplots}
\usepackage{pgfplotstable}
\pgfplotsset{compat=1.18}
\usepgfplotslibrary{fillbetween}
\usetikzlibrary{
  patterns, positioning, shapes, arrows.meta, fit,
  calc, backgrounds, shadows, decorations.pathreplacing,
  decorations.markings, matrix, chains
}

\definecolor{TierLow}{RGB}{220, 50, 47}
\definecolor{TierMed}{RGB}{203, 75, 22}
\definecolor{TierHigh}{RGB}{38, 139, 210}
\definecolor{LogicColor}{RGB}{108, 113, 196}
\definecolor{GridGray}{RGB}{230, 230, 230}
\definecolor{TextGray}{RGB}{80, 80, 80}
\definecolor{C0}{RGB}{0,   63,  92}
\definecolor{C1}{RGB}{47, 120, 177}
\definecolor{C2}{RGB}{255,166,  0}
\definecolor{C3}{RGB}{212, 53,  53}
\definecolor{C4}{RGB}{39, 174, 106}
\definecolor{C5}{RGB}{149, 88, 178}
\definecolor{BgLight}{RGB}{248,249,252}
\definecolor{GridC}{RGB}{215,220,230}
\definecolor{TextD}{RGB}{50,55,65}

\usepackage{setspace}

\begin{document}

\bstctlcite{IEEEexample:BSTcontrol}

\title{HAVE: Host Active Verification Engine for Closing the
       Contextual Reality Gap in Security Digital Twins}
       
\author{Vincenzo Sammartino, Marco Pasquini\\
\textit{Dipartimento di Informatica, Universit\`{a} di Pisa, Pisa, Italia}\\
E-mail: vincenzo.sammartino@phd.unipi.it, m.pasquini10@studenti.unipi.it}

\markboth{IEEE Transactions on Dependable and Secure Computing, 2026}%
{Sammartino and Pasquini: HAVE: Host Active Verification Engine}

\maketitle

\begin{abstract}
Security Digital Twins~(SDTs) provide continuously updated virtual
replicas of infrastructure for threat simulation, yet they rely on
theoretical CVSS scores to assign lateral-movement probabilities---%
creating the \textit{Contextual Reality Gap}: risk is overestimated
where unacknowledged mitigations neutralize exploits, and drastically
underestimated where logic flaws bypass all memory-safety defenses.
We present the \textit{Host Active Verification Engine}~(HAVE), an
SDT extension that deploys a safety-constrained host agent to measure
the empirical probability of compromise~$\hat{p}$ via
maximum-likelihood estimation over snapshot-isolated Bernoulli trials.
A Wilson interval-width confidence weight~$\alpha_w$ propagates~$\hat{p}$
into Monte Carlo simulations via a Bayesian blending rule formally
related to the Beta-Binomial posterior. Evaluation across four
vulnerability classes, three security tiers, and two production binaries shows HAVE reduces~$P_{reach}$ by
$38.2\%$ in false-positive scenarios and increases it by $132.4\%$ in
false-negative scenarios, with a net $+124.1\%$ correction; post-HAVE
estimates vary by only $1.12\times$ across calibration exponents~$\kappa$,
versus $4.6\times$ for CVSS-only baselines.
\end{abstract}

\vspace{-0.2cm}

\section{Introduction}
\label{sec:intro}

\textbf{Context and Motivation.}
The convergence of IT and OT into Cyber-Physical Systems has fundamentally
altered the threat landscape. Security operators require a continuous,
high-fidelity infrastructure risk model~\cite{stouffer2015nist,cheng2017orpheus}.
The Security Digital Twin~(SDT) paradigm addresses this by maintaining
a simulation-capable replica through passive network telemetry~\cite{notline,
fuller2020digital, SammartinoShortPaper}. The passive approach avoids destabilizing legacy
PLC hardware~\cite{muench2018whatyou}, but introduces a critical limitation:
risk quantification relies on CVSS scores~\cite{cvss31, baiardi2026simulation}.

\textbf{The Contextual Reality Gap.}
CVSS provides a \textit{context-free} severity estimate: the same 9.8-rated
stack overflow is empirically near-inert on a PIE+ASLR+Canary binary
but trivially exploitable without those mitigations. Conversely,
a command-injection flaw rated~7.2 remains unconditionally exploitable
irrespective of memory-safety defenses~\cite{szekeres2013sok, baiardi2025securitytwins}. Without
empirical host-level verification, the SDT simultaneously overestimates
mitigated vulnerabilities and underestimates logic flaws.

\textbf{Proposed Solution and Contributions.}
We introduce the \textbf{Host Active Verification Engine}~(HAVE), a
feedback-driven SDT extension that measures host-specific exploitability
empirically via snapshot-isolated Bernoulli trials. Key contributions:
\begin{itemize}
  \item \textbf{Contextual Reality Gap formalization:} first formal
    definition and quantification of CVSS-vs-empirical divergence.
  \item \textbf{Hub-and-Spoke architecture:} cgroups~v2 CPU cap,
    path allow-list confinement, and mTLS agent authentication.
  \item \textbf{Two-phase methodology:} granular static Defense Profile
    plus MLE-based dynamic $\hat{p}$ and TTC estimation.
  \item \textbf{Wilson confidence weighting:} $\alpha_w$ propagates
    $\hat{p}$ via a Bayesian blend formally related to Beta-Binomial.
  \item \textbf{Production binary validation:} CVE-2021-3156 and
    CVE-2021-42013 confirm generalizability.
  \item \textbf{Monte Carlo validation:} five-node dual-path attack
    graph confirms non-linear propagation and $\kappa$-robustness.
\end{itemize}


\vspace{-0.2cm}
\section{Background}
\label{sec:background}

\subsection{Security Digital Twins}
An SDT is a continuously updated, graph-based representation
$\mathcal{G} = (\mathcal{N}, \mathcal{E})$, where each node
$n_i \in \mathcal{N}$ represents a host asset and each directed
edge $(n_i, n_j) \in \mathcal{E}$ represents a potential attack step,
weighted by the probability $p_{ij}$ that an adversary at~$n_i$ can
successfully compromise~$n_j$~\cite{eckhart2018specification,
dietz2020integrating, baiardi2026synthetic}. The SDT continuously ingests telemetry to
update the graph. Simulation engines then traverse~$\mathcal{G}$ to
compute lateral-movement paths, reachability probabilities, and
critical attack vectors. The accuracy of these simulations is directly
bounded by the accuracy of the edge weights~$p_{ij}$, which motivates
the empirical measurement approach of HAVE \cite{Baiardi2026whatif, sammartino2025security}.

\subsection{Limitations of CVSS-Based Risk Scoring}
CVSS v3.1~\cite{cvss31} quantifies vulnerability severity across three
metric groups: Base, Temporal, and Environmental. The Base score
reflects the intrinsic properties of a vulnerability. Temporal metrics
account for exploit code maturity. Environmental metrics allow
per-deployment customization through Modified Impact Metrics, and if
correctly populated they partially account for host-specific
mitigations. However, maintaining accurate per-host Environmental
profiles incurs significant operational overhead, and in practice this
sub-score is rarely populated~\cite{cvss31, baiardi2024anticipating}.

Two structural limitations make CVSS inadequate as a sole risk source
for an SDT. First, the Base score is \textit{context-free}: it measures
the worst-case scenario severity without accounting for host-specific
mitigations. Second, CVSS does not distinguish between vulnerability
classes that interact differently with mitigation mechanisms. A
memory-corruption flaw and a command-injection flaw may share a
similar CVSS score but have entirely different exploitability profiles
depending on the presence of memory-safety features~\cite{bozorgi2010beyond}.
The Exploit Prediction Scoring System~(EPSS)~\cite{jacobs2021epss}
partially addresses exploitability prediction using external threat
intelligence signals, but it remains a population-level statistical
model that does not reflect the specific configuration of an
individual host. HAVE addresses this gap directly by automating the
measurement of host-specific exploitability, yielding data equivalent
to a fully populated Environmental Score without the manual profiling
burden.

\subsection{Host-Level Security Primitives}
Modern Linux systems layer four complementary defenses against
memory-corruption attacks~\cite{szekeres2013sok}.
\textbf{ASLR} randomizes stack, heap, and library base addresses at
load time; its effectiveness requires PIE compilation, since a
non-PIE binary loads its code segment at a fixed address regardless
of kernel ASLR policy~\cite{shacham2004effectiveness}.
\textbf{PIE} and \textbf{Full RELRO} enable code-segment
randomization and make the Global Offset Table read-only after
dynamic linking, respectively.
\textbf{Stack Canaries}~\cite{cowan1998stackguard} detect
stack-frame corruption before return-address overwrite but offer no
protection against heap-segment operations.
\textbf{NX/DEP} prevents direct shellcode injection; attackers
circumvent it via Return-Oriented Programming~\cite{shacham2007geometry}.
The interaction between these primitives --- not their individual
presence --- determines actual exploitability, as formally captured
by $R_{eff}$ in Section~\ref{sec:method:static}.

\section{Related Work}
\label{sec:relwork}

\subsection{Digital Twin Frameworks for Cyber-Resilience}
Fuller et al.~\cite{fuller2020digital} provide a comprehensive survey
of digital twin enabling technologies and identify security simulation
as a primary application domain. Eckhart and
Ekelhart~\cite{eckhart2018specification} demonstrate a
specification-based state-replication approach for CPS, establishing
the principle that a twin should mirror runtime behavior rather than
static configurations. Dietz and Pernul~\cite{dietz2020integrating}
integrate SDT simulations with Security Operations Centers, showing
how automated attack playbooks can feed twin state updates. The
NotLine framework~\cite{notline}, which HAVE extends, builds
high-fidelity network-layer twins through passive traffic analysis,
avoiding destabilizing probes while maintaining topological accuracy.
Baiardi et al.~\cite{baiardi2025ai} augment the NotLine topology
with Graph Neural Network-based risk inference; however, their
approach still ultimately relies on CVSS-derived prior weights,
a limitation directly addressed by HAVE's empirical measurements.

\subsection{Network-Level Vulnerability Assessment}
Active network scanners such as Nmap~\cite{lyon2009nmap} and
OpenVAS~\cite{greenbone2009openvas} detect vulnerabilities by probing
exposed services and comparing banners against CVE signatures. These
tools operate externally and cannot directly observe the host's
internal security configuration. Banner-based fingerprinting is known
to produce significant false-positive rates when service versions are
not precisely identifiable~\cite{moberg2014configuration}. More
critically, external scanning is architecturally incompatible with
OT environments: sending unexpected traffic to a programmable logic
controller can trigger safety faults. HAVE avoids this by deploying
an internal agent that never generates external network probes.

\subsection{Host-Based Auditing and Configuration Management}
Agent-based auditing tools such as OSSEC and Wazuh maintain real-time
inventories of installed packages and system configurations, alerting
on deviations from a security baseline. While valuable, these tools
perform \textit{static} inventory rather than dynamic exploitability
testing. They report the presence of a vulnerable library but cannot
determine whether the vulnerable code path is reachable or whether
existing mitigations render the vulnerability practically inert.
HAVE's static phase is conceptually analogous to host-based auditing
but is explicitly designed as the precursor to a dynamic phase that
provides the missing exploitability verdict~\cite{bozorgi2010beyond}.

\subsection{Automated Exploit Generation}
Avgerinos et al.~\cite{avgerinos2011aeg,avgerinos2014automatic}
pioneered Automated Exploit Generation~(AEG), demonstrating that
symbolic execution can automatically generate working exploits for
programs with known vulnerability classes. Shoshitaishvili et
al.~\cite{shoshitaishvili2016sok} provide a comprehensive taxonomy of
binary analysis techniques applicable to exploitation. The critical
distinction between AEG and HAVE lies in their objective: AEG
\textit{discovers} new exploitable conditions, whereas HAVE uses a
curated library of pre-validated exploits to \textit{measure the
efficacy of defenses} against known vulnerability classes. This focus
on ``mitigation verification'' makes HAVE applicable in production
risk assessment contexts where the vulnerability is already known and
the unknown quantity is its practical exploitability given the target
host's configuration.

\subsection{Risk Quantification in Attack Graphs}
Attack graph formalisms~\cite{sheyner2002automated, ou2006scalable}
model multi-step attack paths as directed graphs. Bayesian attack
graphs~\cite{frigault2008measuring, poolsappasit2012dynamic} extend
this formalism by associating conditional probabilities with each
attack step. McQueen et al.~\cite{mcqueen2006time} introduce the
Mean Time-to-Compromise metric as a complementary measure of attacker
effort. Homer et al.~\cite{homer2013aggregating} demonstrate methods
for aggregating per-step vulnerability metrics into infrastructure-wide
security scores. HAVE directly addresses the data-quality problem
identified across this body of work: it replaces CVSS-derived,
context-free edge weights with empirically measured, host-specific
exploitability probabilities.

\subsection{Active Testing Safety in OT Environments}
The risks of active security testing in CPS environments are well
documented~\cite{stouffer2015nist, corteggiani2018inception,
muench2018whatyou}. Muench et al.~\cite{muench2018whatyou} demonstrate
that software corruption in embedded systems does not manifest in
observable crashes with the same reliability as in desktop systems.
Corteggiani et al.~\cite{corteggiani2018inception} propose an
emulation-based approach for security testing of embedded firmware.
Our work is aligned with this philosophy: by mandating VM-based
sandboxing in OT contexts and enforcing hard CPU caps via
\texttt{cgroups}~\cite{menage2007cgroups}, HAVE ensures that active
verification cannot propagate instability to production hardware.

\subsection{Breach-and-Attack Simulation Platforms}
\label{sec:relwork:bas}
Commercial Breach-and-Attack Simulation~(BAS) platforms --- including
SafeBreach, AttackIQ, and XM~Cyber --- represent the closest class of
prior work to HAVE's operational objective of automated, agent-based
exploit execution for security measurement. These platforms deploy
agents that execute pre-defined attack scenarios against live
infrastructure to measure whether security controls block or permit
specific Tactics, Techniques, and Procedures~(TTPs) as catalogued
by MITRE ATT\&CK~\cite{dulaunoy2020mitre}.

HAVE is architecturally distinguished from this class of platforms
along three fundamental dimensions. \textit{First}, BAS platforms
operate at the campaign and TTP level: they report pass/fail
outcomes for discrete attack scenarios but do not produce
statistically calibrated probability estimates over~$N$ I.I.D.\ trials
that can be propagated as edge weights into a probabilistic attack
graph. The Wilson confidence interval and the $\alpha_w$ blending
rule (Section~\ref{sec:method:monte}) are methodologically absent from
BAS platforms. \textit{Second}, BAS platforms are not designed for
integration with SDT risk models: their outputs are security control
assessment reports, not parameterized edge-weight updates for Monte
Carlo simulation engines. The feedback loop from HAVE's
empirical~$\hat{p}$ to the SDT's~$P_{reach}(n^*)$ is the architectural
contribution absent in the BAS class. \textit{Third}, BAS platforms
do not address the OT safety constraints that are central to HAVE's
design: none of the platforms surveyed in the literature expose
formal resource-governance interfaces~(cgroups~CPU capping), directory-based
authorization allow-lists, or snapshot-isolation guarantees that
prevent active testing from destabilizing production PLC firmware.
In summary, BAS platforms and HAVE address complementary but
non-overlapping portions of the security assessment problem: BAS
provides breadth~(TTP coverage across the kill chain), while HAVE
provides depth~(statistically rigorous, per-vulnerability,
per-host exploitability quantification for SDT integration).

\section{System and Threat Model}
\label{sec:model}

\subsection{System Model}
We consider an enterprise or industrial infrastructure modeled as a
graph~$\mathcal{G} = (\mathcal{N}, \mathcal{E})$ continuously
maintained by an SDT. The node set
$\mathcal{N} = \mathcal{N}_{IT} \cup \mathcal{N}_{OT}$ includes both
standard IT assets (servers, workstations, routers) and OT assets
(PLCs, HMIs, SCADA servers). Each node~$n_i$ has an associated
security configuration~$\sigma_i$ comprising the kernel ASLR state
and the binary-level feature vector~$\mathbf{v}_{bin}$ as defined
in Section~\ref{sec:method:static}. The passive monitoring layer
continuously populates~$\mathcal{G}$ by observing network telemetry.
The HAVE Controller, embedded in the SDT's analytics layer, augments
this topological data with empirical exploitability data derived from
host-level verification campaigns.

\subsection{Adversary Model}
We model a threat actor consistent with the MITRE ATT\&CK
framework~\cite{dulaunoy2020mitre}:

\textbf{Capability.} The adversary has initial access to at least
one compromised node~$n_0 \in \mathcal{N}$ and seeks to perform
lateral movement to reach high-value targets. The adversary possesses
a library of known exploits for public CVEs but does not have access
to zero-day vulnerabilities. The adversary is resource-rational:
given multiple potential attack paths, the adversary will select the
path of minimum expected effort, quantified by Time-to-Compromise.

\textbf{Constraints.} The adversary cannot modify the SDT's model or
inject false telemetry (the mTLS authentication model prevents
spoofing, as detailed in Section~\ref{sec:arch}). The adversary is
not assumed to have physical access to the hardware.

\textbf{Zero-Day Scope.} HAVE's measurements cover the known-CVE
attack surface constituting the primary threat against most
deployments. Operators in critical infrastructure contexts should
supplement HAVE with threat-intelligence feeds providing early
warning of emerging zero-day campaigns.

\subsection{HAVE Safety Model}
The introduction of active code execution into a production
infrastructure carries inherent risks that must be addressed by the
verification framework itself. We define three safety requirements:

\noindent\textbf{SR-1~(Resource Boundedness).} The verification process must
consume a bounded fraction of the target host's CPU resources,
preventing any Denial-of-Service~(DoS) condition.

\noindent\textbf{SR-2~(Scope Confinement).} The verification agent must only
execute payloads against explicitly authorized targets, preventing
repurposing as a lateral-movement tool.

\noindent\textbf{SR-3~(State Integrity).} The verification process must not
leave residual artifacts or system corruption on the target host after
each trial.

These requirements are formalized and enforced by the three safety
mechanisms detailed in Section~\ref{sec:arch}.

\section{HAVE Architecture}
\label{sec:arch}

The Host Active Verification Engine is engineered as a feedback-driven
extension of the SDT ecosystem. Rather than operating as a standalone
scanner, HAVE is a tightly integrated component that receives triggers
from the passive monitoring layer and returns empirical measurements
that directly update the twin's risk model.

\subsection{Design Principles}
The architecture is governed by three principles derived from the
safety model of Section~\ref{sec:model}: (i)~\textit{Safety by
Design}, ensuring every code path within the agent is constrained by
the three safety requirements; (ii)~\textit{Minimal Footprint},
deploying the lightest possible privileged process on the target host;
(iii)~\textit{Event-Driven Activation}, triggering verification
campaigns only in response to meaningful state changes detected by the
passive layer (e.g., a new CVE published for a running service, or
a configuration drift detected via telemetry), thereby minimizing
unnecessary overhead.

\subsection{Hub-and-Spoke Topology}
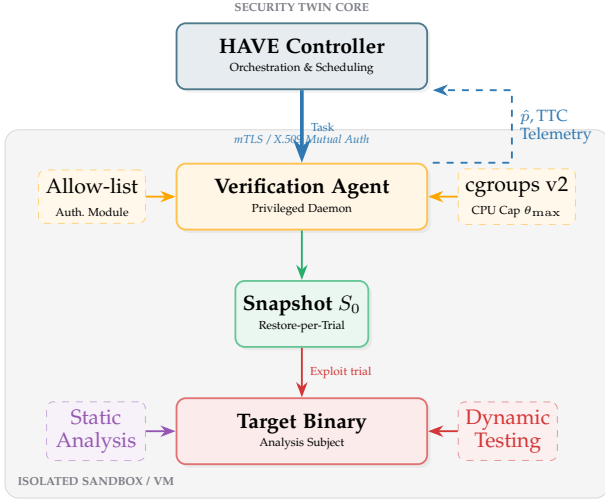
\begin{figure}[t!]
\centering
\resizebox{0.9\columnwidth}{!}{%
\begin{tikzpicture}[
  >=Stealth,
  font=\small,
  box/.style={
    rectangle, rounded corners=4pt, draw=#1!70, thick,
    fill=#1!10, minimum height=1.0cm, align=center,
    drop shadow={opacity=0.10, shadow xshift=1.5pt, shadow yshift=-1.5pt}
  },
  sbox/.style={
    rectangle, rounded corners=3pt, draw=#1!50, thin,
    fill=#1!8, minimum height=0.8cm, align=center, dashed
  },
  lbl/.style={font=\tiny\bfseries, color=TextD!65},
  arr/.style={->, thick, color=TextD!70},
  rarr/.style={->, thick, dashed, color=C1, line width=0.8pt}
]

\node[box=C0, minimum width=3.8cm] (ctrl)
  {\textbf{HAVE Controller}\\[-2pt]{\tiny Orchestration \& Scheduling}};
\node[above=0.05cm of ctrl, lbl] {SECURITY TWIN CORE};

\node[below=0.55cm of ctrl, font=\tiny\itshape, color=C1] (mtls)
  {mTLS / X.509 Mutual Auth};

\node[box=C2, minimum width=3.8cm, below=1.1cm of ctrl] (agent)
  {\textbf{Verification Agent}\\[-2pt]{\tiny Privileged Daemon}};

\node[sbox=C2, minimum width=1.5cm, left=0.45cm of agent] (auth)
  {Allow-list\\{\tiny Auth.\ Module}};
\node[sbox=C2, minimum width=1.5cm, right=0.45cm of agent] (cpu)
  {cgroups v2\\{\tiny CPU Cap $\theta_{\max}$}};

\node[box=C4, minimum width=2.0cm, below=0.75cm of agent] (snap)
  {\textbf{Snapshot} $S_0$\\[-2pt]{\tiny Restore-per-Trial}};

\node[box=C3, minimum width=3.8cm, below=0.75cm of snap] (bin)
  {\textbf{Target Binary}\\[-2pt]{\tiny Analysis Subject}};

\node[sbox=C5, minimum width=1.5cm, left=0.45cm of bin, text=C5] (stat)
  {Static\\Analysis};
\node[sbox=C3, minimum width=1.5cm, right=0.45cm of bin, text=C3] (dyn)
  {Dynamic\\Testing};

\begin{scope}[on background layer]
  \node[
    fit=(auth)(cpu)(agent)(snap)(bin)(stat)(dyn),
    fill=black!4, rounded corners=6pt,
    draw=black!15, inner sep=0.5cm
  ] (sandbox) {};
\end{scope}
\node[above right=0.08cm and 0.08cm of sandbox.south west, lbl]
  {ISOLATED SANDBOX / VM};

\draw[arr, C1, ultra thick] (ctrl) -- (agent)
  node[midway, right, font=\tiny] {Task};
\draw[arr, C2] (auth) -- (agent);
\draw[arr, C2] (cpu)  -- (agent);
\draw[arr, C4] (agent) -- (snap);
\draw[arr, C3] (snap)  -- (bin)
  node[midway, right, font=\tiny] {Exploit trial};
\draw[arr, C5] (stat) -- (bin);
\draw[arr, C3] (dyn)  -- (bin);

\draw[rarr] ([xshift=2pt]agent.north east)
  -- ++(1.2,0) |- node[pos=0.25, right, font=\scriptsize, align=left, C1]
  {$\hat{p}$,\,TTC\\Telemetry}
  ([xshift=2pt]ctrl.south east);

\end{tikzpicture}}
\caption{\textbf{HAVE System Architecture (Hub-and-Spoke).}
The Controller~(navy) dispatches verification tasks over an mTLS
channel. The Agent~(amber) enforces an allow-list and a
\texttt{cgroups}~v2 CPU cap before interacting with the target binary.
Each trial is preceded by a hypervisor snapshot restore~(green) to
guarantee I.I.D.\ trial conditions. Empirical telemetry~($\hat{p}$, TTC)
flows back to update the SDT's risk model.}
\label{fig:architecture}
\end{figure}

The system is organized in a Hub-and-Spoke topology
(Fig.~\ref{fig:architecture}). The \textit{Orchestration Controller}
(the Hub) resides within the SDT's analytics layer. The
\textit{Verification Agents} (the Spokes) are deployed on or near
the target assets. All communication traverses a mutually authenticated
TLS channel.

\subsection{The Orchestration Controller}
The Controller translates high-level risk queries into discrete,
atomic verification tasks. It maintains a dynamic registry of all
active Verification Agents and manages campaign scheduling. The
Controller operates in event-driven mode: verification campaigns are
triggered by~(i) detection of a new vulnerable service via passive
telemetry, (ii) publication of a CVE affecting a catalogued asset,
or~(iii) an explicit operator request. This design avoids continuous,
background exploitation attempts that would consume resources
unnecessarily. Upon receiving results from an Agent, the Controller
applies the Bayesian update rule~(Section~\ref{sec:method:monte}) to
refine the edge weight~$p_{ij}$ in~$\mathcal{G}$, and propagates the
updated simulation to the SDT's risk dashboard. The Controller also
maintains an audit log of all verification campaigns for compliance
and forensic purposes.

\subsection{The Verification Agent: A Safety-Constrained Daemon}
The Verification Agent is a lightweight privileged daemon. In IT
contexts it may run directly on the host OS; in OT contexts, it is
mandated to execute within a high-fidelity VM sandbox mirroring the
production environment, ensuring that even a catastrophic verification
failure cannot destabilize the physical hardware.

The agent enforces three immutable safety constraints that directly
satisfy requirements SR-1 through SR-3.

\subsubsection{Resource Governance: CPU Capping~(SR-1)}
The resource governance constraint is implemented via Linux
\texttt{cgroups}~v2~\cite{menage2007cgroups}, which operates through
the Completely Fair Scheduler~(CFS) over discrete periods. Let
$\Delta t_{CFS}$ denote the CFS quota period (default 100\,ms). For
each scheduler period~$k \in \mathbb{Z}_{\geq 0}$, let $T_{exec}^{(k)}$
denote the agent's CPU execution time within that period. The
discrete-time resource governance constraint is:
\begin{equation}
  \forall k \in \mathbb{Z}_{\geq 0}:\quad
  \frac{T_{exec}^{(k)}}{\Delta t_{CFS}} \;\leq\; \theta_{\max},
  \label{eq:cgroup}
\end{equation}
where $\theta_{\max} \in (0,1]$ is the maximum allowable CPU utilization
fraction defined in the local configuration manifest. This constraint
is enforced by writing
$\bigl\lfloor\theta_{\max} \cdot \Delta t_{CFS} \cdot 10^6\bigr\rfloor$
to the agent's \texttt{cgroup}~v2 leaf node at
\texttt{/sys/fs/cgroup/have.slice/cpu.max} (e.g., for
$\theta_{\max}=0.25$ and $\Delta t_{CFS}=100\,\text{ms}$: value
``\texttt{25000~100000}''). The kernel scheduler throttles the agent
process for the remainder of any period in which Eq.~\eqref{eq:cgroup}
is violated. A typical OT deployment sets $\theta_{\max} = 0.15$,
reserving 85\% of CPU capacity for critical control logic.

\subsubsection{Authorization Confinement: Path Allow-list~(SR-2)}
Let $\mathcal{D}_{allow} = \{d_1, d_2, \dots, d_n\}$ be the set of
directory paths explicitly authorized in the agent's root-of-trust
configuration. Given a verification request targeting a binary at
path~$p_{target}$, the agent applies a validation function:
\begin{equation}
V(p_{target}) = \begin{cases}
\text{ACCEPT} & \text{if } \exists\, d_i \in \mathcal{D}_{allow} :
                p_{target} \subseteq d_i \\
\text{REJECT} & \text{otherwise.}
\end{cases}
\label{eq:allowlist}
\end{equation}
Prior to this check, $p_{target}$ is sanitized by resolving symbolic
links and stripping directory-traversal sequences~(\texttt{../}).
Requests targeting system paths outside~$\mathcal{D}_{allow}$ are
summarily rejected and logged as security violations. This design
mitigates the \textit{confused deputy} attack, where a compromised
Controller attempts to coerce the Agent into executing payloads on
arbitrary system binaries~\cite{rose2020zerotrust}.

\subsubsection{Cryptographic Identity: Mutual TLS~(Security)}
The command-and-control channel between Controller~$C$ and Agent~$A$
is secured via mutually authenticated TLS~(mTLS). Both entities must
present valid X.509 certificates signed by an internal Certificate
Authority~$\mathit{CA}_{int}$:
\begin{equation}
\mathrm{Verify}(Cert_C,\, \mathit{CA}_{int}) \;\land\;
\mathrm{Verify}(Cert_A,\, \mathit{CA}_{int})
\;\Rightarrow\; \mathrm{Session}(C,\, A).
\label{eq:mtls}
\end{equation}
This bidirectional authentication ensures that~(i) the Agent accepts
commands only from a verifiable Controller, and~(ii) the Controller
ingests data only from verifiable Agents, preventing pollution of the
SDT's risk model with spoofed telemetry.

\subsection{Security Twin Integration Loop}
Upon completion of a verification campaign, the Controller updates
the risk model as follows. Let $e_{ij} \in \mathcal{E}$ be the attack
step from~$n_i$ to~$n_j$ whose weight was previously set to
$p_{ij}^{(0)}$ derived from the CVSS score. The Bayesian blending
rule~(Eq.~\eqref{eq:bayes}) governs the weight assigned to the
empirical estimate relative to the CVSS prior as a function of the
Wilson confidence interval width. The updated graph~$\mathcal{G}'$
is then provided to the Monte Carlo simulation engine.

\section{Analysis Methodology}
\label{sec:method}

The methodological core of HAVE is predicated on the assertion that
risk is a dynamic function of the execution environment, not a static
property of the vulnerable code. HAVE implements a two-phase pipeline
that sequentially evaluates the theoretical defense posture~(Phase~I:
Static Analysis) and the empirical resistance~(Phase~II: Dynamic
Analysis).

\subsection{Phase~I: Granular Static Analysis}
\label{sec:method:static}

The objective of Phase~I is to construct a high-resolution
\textit{Defense Profile}~$\Pi_i$ for the target binary, providing
both a theoretical risk ceiling and the input parameters that configure
the subsequent dynamic phase.

\textbf{Kernel Configuration.}
The Agent reads the kernel ASLR policy from
\texttt{/proc/sys/kernel/randomize\_va\_space}, obtaining
$\alpha_{kern} \in \{0, 1, 2\}$.

\textbf{Binary-Level Feature Extraction.}
Using the \texttt{checksec}~\cite{checksec} functionality of
\texttt{pwntools}~\cite{pwntools}, the Agent parses the target binary's
ELF headers to extract the binary feature vector:
\begin{equation}
  \mathbf{v}_{bin} \;=\;
  \langle \beta_{pie},\; \gamma_{can},\; \delta_{nx},\;
          \epsilon_{relro} \rangle,
  \label{eq:featvec}
\end{equation}
where $\beta_{pie}, \gamma_{can}, \delta_{nx} \in \{0,1\}$ indicate
PIE, Stack Canary, and Non-Executable memory status, respectively,
and $\epsilon_{relro} \in \{0, \text{Partial}, \text{Full}\}$
indicates the RELRO level. The Agent additionally scans the binary's
symbol table for \texttt{\_FORTIFY\_SOURCE} instrumentation.

\textbf{Effective Randomization.}
A critical methodological point is that kernel ASLR and binary PIE
are not independent: their interaction determines the actual degree
of address-space randomization. We define the Effective Randomization:
\begin{equation}
R_{eff} \;=\; f(\alpha_{kern},\, \beta_{pie}) \;=\;
\begin{cases}
\textit{High}   & \text{if } \alpha_{kern} \geq 1 \;\land\;
                  \beta_{pie} = 1 \\
\textit{Low}    & \text{otherwise.}
\end{cases}
\label{eq:reff}
\end{equation}
A kernel ASLR setting of~2 with a non-PIE binary ($\beta_{pie} = 0$)
yields $R_{eff} = \textit{Low}$, since the code segment loads at a
deterministic address. This interaction is precisely the nuance that
CVSS and static inventory tools fail to capture.

The complete Defense Profile is:
\begin{equation}
  \Pi_i \;=\; \bigl\{ \alpha_{kern},\; \mathbf{v}_{bin},\;
               R_{eff},\; \textit{FORTIFY} \bigr\}.
  \label{eq:defprofile}
\end{equation}

\subsection{Phase~II: Empirical Dynamic Analysis}
\label{sec:method:dynamic}

Phase~II subjects the target binary to a statistically rigorous
sequence of controlled exploit trials.

\textbf{I.I.D.\ Trial Design.}
Dynamic exploitation is inherently destabilizing. A failed exploit
may leave the process in an undefined state. To ensure that each
trial is drawn from the same distribution, we model the verification
sequence as a set of I.I.D.\ Bernoulli trials
$\mathcal{T} = \{t_1, t_2, \dots, t_N\}$ and enforce independence
via a \textit{Restore-per-Trial} protocol.

Let $S_0$ be the \textit{ReadyState} snapshot of the VM, captured
immediately after the target binary is loaded and ready for
exploitation. The execution flow for each trial~$t_i$ is:
\begin{equation}
  \text{Execute}(t_i) \;\leftarrow\;
  \text{Restore}(S_0) \;\circ\; \text{RunExploit}(\texttt{payload}).
  \label{eq:trial}
\end{equation}
This guarantees that system entropy, memory layout, and file
descriptors are reset to an identical baseline before every trial,
satisfying~SR-3.

\textbf{Metric Derivation.}
Each trial~$t_i$ yields a binary outcome~$x_i \in \{0, 1\}$, where
$x_i = 1$ denotes successful compromise (verified by matching an
expected exit code, e.g., spawning a controlled shell). Upon
completion of~$N$ trials, the Empirical Probability of
Compromise~$\hat{p}$ is computed via Maximum Likelihood Estimation
for a Binomial model. Let $S = \sum_{i=1}^{N} x_i$:
\begin{equation}
  \hat{p} \;=\; \frac{S}{N}.
  \label{eq:phat}
\end{equation}
A 95\% Wilson score confidence interval
$[\hat{p}_L, \hat{p}_U]$ accompanies~$\hat{p}$:
\begin{equation}
  \hat{p}_{L,U} \;=\; \frac{\hat{p} + \tfrac{z^2}{2N} \mp
  z\sqrt{\tfrac{\hat{p}(1-\hat{p})}{N} + \tfrac{z^2}{4N^2}}}%
  {1 + \tfrac{z^2}{N}},
  \label{eq:wilson}
\end{equation}
where $z = 1.96$ for a 95\% confidence level. Additionally, the
Time-to-Compromise is computed over successful trials
$K = \{i \mid x_i = 1\}$:
\begin{equation}
  TTC \;=\; \frac{1}{|K|} \sum_{i \in K} \tau_i,
  \label{eq:ttc}
\end{equation}
where~$\tau_i$ is the wall-clock execution latency of trial~$t_i$.
Following McQueen et al.~\cite{mcqueen2006time}, TTC serves as a
proxy for attacker effort.

\subsection{Risk Model Integration: Monte Carlo Simulation}
\label{sec:method:monte}

The SDT's simulation engine models lateral movement via Monte Carlo
traversal of~$\mathcal{G}$. In each simulation run, each edge~$e_{ij}$
is sampled as a Bernoulli trial with success probability~$p_{ij}$.
After $M$ simulation runs (typically $M = 10{,}000$), the fraction of
runs in which the adversary reaches a target node~$n^*$ estimates
the end-to-end compromise probability~$P_{reach}(n^*)$.

\subsubsection{CVSS-to-Probability Mapping and Calibration Exponent}
\label{sec:method:kappa}
Prior to HAVE integration, edge weights are initialized from CVSS
scores via the monotone power mapping:
\begin{equation}
  p_{ij}^{(0)} \;=\; \Bigl(\frac{\text{CVSS}_{ij}}{10}\Bigr)^{\!\kappa},
  \label{eq:cvssmap}
\end{equation}
where $\kappa > 0$ is a calibration exponent governing the concavity
of the mapping. We adopt $\kappa = 1$ as the baseline parameterization,
yielding a linear mapping~$p_{ij}^{(0)} = \text{CVSS}_{ij}/10$. This
choice is consistent with prior work~\cite{poolsappasit2012dynamic} and
has a natural interpretation: the CVSS score normalized to~$[0,1]$
serves directly as the prior probability of exploitation. We note that
$\kappa < 1$ compresses mid-range scores toward~1 (i.e., more
pessimistic prior), while $\kappa > 1$ expands them toward~0 (more
optimistic prior). A sensitivity analysis over
$\kappa \in \{0.5, 1.0, 1.5, 2.0\}$ is presented in
Section~\ref{sec:results:kappa}, demonstrating that post-HAVE risk
estimates are robust to this parameterization choice, while the
CVSS-only baseline varies substantially.

\subsubsection{Bayesian Blending Rule}
HAVE replaces the CVSS prior with the empirical posterior:
\begin{equation}
  p_{ij} \;\leftarrow\; \alpha_w \cdot \hat{p} \;+\;
  (1 - \alpha_w) \cdot p_{ij}^{(0)},
  \label{eq:bayes}
\end{equation}
where $\alpha_w \in [0,1]$ is a confidence weight defined as a
monotone-decreasing function of the Wilson interval width:
\begin{equation}
  \alpha_w \;=\; 1 \;-\; \bigl(\hat{p}_U - \hat{p}_L\bigr).
  \label{eq:alphaconf}
\end{equation}
As $N\to\infty$, the Wilson width vanishes and $\alpha_w\to 1$,
so the posterior converges to the empirical estimate; for small~$N$,
$\alpha_w$ defaults toward the CVSS prior.

\subsubsection{Formal Relationship to the Beta-Binomial Posterior}
\label{sec:method:betabin}
The Bayesian-canonical treatment assigns a Beta prior
$p_{ij}\!\sim\!\mathrm{Beta}(\alpha_0,\beta_0)$ with
$\alpha_0 = p_{ij}^{(0)}\cdot n_0$ and
$\beta_0 = (1-p_{ij}^{(0)})\cdot n_0$, yielding a posterior mean of
the same convex-combination form as Eq.~\eqref{eq:bayes} with weight
$\alpha_w^{BB}=N/(n_0+N)$.
The critical distinction is that $\alpha_w^{BB}$ depends only on~$N$
for fixed~$n_0$, whereas the Wilson-width weight~$\alpha_w$ is
\emph{state-dependent}: it depends jointly on~$N$ and~$\hat{p}$.
This is epistemically superior for risk assessment: when HAVE observes
extreme outcomes ($\hat{p}\in\{0,1\}$), the Wilson interval is narrow
even at moderate~$N$, correctly assigning high confidence to the
empirical estimate, while a uniform 50/50 outcome at the same~$N$
produces a wide interval and appropriately discounts the prior.
In all experiments reported here $N\geq 30$, bounding the maximum
Wilson width to~$0.364$ and ensuring $\alpha_w\geq 0.636$ throughout.

\section{Implementation}
\label{sec:impl}

\subsection{Agent Implementation}
The Verification Agent is implemented as a Python daemon
communicating with the Controller over gRPC with mutual TLS. The
exploit execution module uses \texttt{pwntools}~\cite{pwntools} as
the primary binary interaction framework. The static analysis module
invokes \texttt{checksec}~\cite{checksec} programmatically via
subprocess and parses the JSON output into a structured Defense
Profile~$\Pi_i$. Hypervisor snapshot management is implemented via
the \texttt{libvirt} API, supporting QEMU/KVM and VMware ESXi backends.
The \texttt{cgroups}~v2 CPU quota is written to
\texttt{/sys/fs/cgroup/have.slice/cpu.max} at agent initialization.

\subsection{Exploit Library and Self-Test Harness}
\label{sec:impl:library}
The dynamic testing module is supplied by a curated exploit library
implementing benign payloads for each supported vulnerability class.
A \textit{benign payload} demonstrates a successful compromise
(e.g., achieving arbitrary code execution and writing a controlled
flag value to a predetermined memory address) without performing any
destructive, persistent, or network-propagating action.

The library currently comprises 12 exploit modules~(3 per vulnerability
class, one per security tier), totaling approximately 2,400 lines of
Python. Each module is parameterized by the Defense Profile~$\Pi_i$,
selecting among exploitation strategies at runtime based on~$R_{eff}$
and~$\mathbf{v}_{bin}$.

Each module includes a \textit{self-test harness} that validates
correct exploit behavior on a purpose-built reference environment
before any production deployment. The self-test procedure proceeds
as follows: (i)~the module is executed against a reference VM~(Ubuntu
22.04, same kernel version and security tier as the target
configuration) whose ground-truth exploitability is known a priori
from the vulnerability class design; (ii)~the harness verifies that
the benign payload writes the expected 8-byte flag to a predetermined
address in the BSS segment; (iii)~the harness confirms that the
target process terminates cleanly~(exit code~0) and that no residual
file-system artifacts are created; and~(iv)~for stochastic exploits,
the harness additionally validates that the observed success rate over
a reference run of~$N_{\text{ref}} = 30$ trials falls within the
expected range for the tier~(e.g., $\hat{p} = 0.00$ for a
fully-protected High-tier binary, or~$\hat{p} > 0.90$ for a
Low-tier binary). Self-test failures block module deployment and
raise an alert via the Controller audit log. Library versioning
follows semantic versioning: new vulnerability classes increment the
MINOR version; new exploitation techniques increment PATCH.

An important scope boundary is the following: the exploit library
measures the exploitability of a vulnerability class using
\textit{specific, implemented exploit techniques} rather than the
exploitability of the vulnerability class in the abstract. The
measured~$\hat{p}$ is therefore bounded above by the quality and
completeness of the library's exploit implementation for the specific
tier. For example, the format-string High-tier estimate of
$\hat{p} = 0.33$ reflects the efficacy of the specific
address-leakage strategy implemented, not a universal ceiling on all
possible format-string exploits. Operators should treat~$\hat{p}$ as
a \textit{lower bound on exploitability under the modeled adversary
capability level}, consistent with the threat model of
Section~\ref{sec:model}.

\subsection{Testbed Configuration}
All experiments were conducted on an Ubuntu~22.04~LTS host (kernel
5.15, x86-64) with 8 CPU cores and 16\,GB RAM, running the
Verification Agent inside a QEMU/KVM virtual machine allocated 2
vCPUs and 2\,GB RAM. The CPU cap was set to $\theta_{\max} = 0.25$
during verification campaigns. A ReadyState snapshot~$S_0$ was
created for each vulnerability tier immediately after binary loading,
and restored via \texttt{libvirt} API before every trial. The mean
snapshot restore latency was $98 \pm 12\,\text{ms}$.

It is important to contextualize this latency correctly. The $98\,\text{ms}$
restore latency affects the \textit{duration of the overall verification
campaign} (i.e., the calendar time required to complete~$N$ trials for
a given vulnerability), not the PLC control loop. Since verification
campaigns are triggered by CVE publication events~(which occur at most
on the order of days) and are executed against VM replicas rather than
the physical hardware, the latency has no operational impact on
real-time PLC control. Standard PLC scan cycles of 1--100\,ms apply to
the physical controller; the VM replica's timing fidelity is relevant
only to the interpretation of TTC values, which we explicitly qualify
as relative ordering metrics~(Section~\ref{sec:discussion:deployment}).

\subsection{Computational Overhead Characterization}
\label{sec:impl:overhead}

To confirm enforcement correctness, CPU utilization was measured via
\texttt{perf stat} at 1\,Hz across all campaigns. Mean utilization
ranged from 5.3\% (Logic Flaw) to 16.5\% (Format String, High tier);
the highest instantaneous value observed was 23.4\%, providing a
6.6-point margin below the $\theta_{\max}=0.25$ cap. Memory footprint
remained below 170\,MB (dominated by \texttt{pwntools} process space);
read IOPS were below 60, driven by snapshot-restore operations.

\subsection{Production Binary Validation}
\label{sec:impl:production}

To validate that the mitigation-effectiveness patterns identified in
the synthetic corpus generalize to real-world software, we conducted
additional experiments on two production binaries with registered CVE
identifiers. These were selected to match the two most operationally
significant findings: the complete neutralization of memory-corruption
exploits by full PIE+ASLR~(heap class), and the unconditional
exploitability of logic flaws across all security tiers \cite{BaiardiSammartino2025Validation}.

\textbf{CVE-2021-3156 (sudo ``Baron~Samedit'').}
CVE-2021-3156 is a heap-based buffer overflow
in the \texttt{sudoedit} command of sudo~$\leq$\,1.9.5p1, disclosed
in January~2021 with a public proof-of-concept exploit.
The flaw arises from improper handling of a trailing backslash in
command-line argument parsing, leading to a heap buffer overflow that
can be leveraged to achieve privilege escalation to root. We installed
sudo~1.8.31 from the Ubuntu~20.04 LTS package repositories (CVSS
Base Score: 7.8). Experiments were conducted at two protection levels
aligned with our tier definitions:

\begin{itemize}
  \item \textbf{Low-equivalent} (sudo compiled without PIE; ASLR
    disabled via \texttt{randomize\_va\_space=0}): The public
    proof-of-concept exploit succeeded with
    $\hat{p} = 1.00$~($N = 50$, Wilson CI~$[0.929, 1.000]$,
    TTC $= 22.1 \pm 3.4\,\text{ms}$). The higher TTC relative to
    the synthetic heap Low-tier~(15.8\,ms) reflects process-fork
    overhead inherent to the sudo architecture.

  \item \textbf{High-equivalent} (Ubuntu~20.04 default: PIE-enabled
    sudo binary; ASLR = 2): The exploit failed with
    $\hat{p} = 0.00$~($N = 100$, Wilson upper bound~$0.037$).
    The exploit relies on predictable heap-chunk offsets derived
    from a non-randomized heap base address; full PIE randomizes
    these offsets, invalidating the hardcoded chunk distances.
\end{itemize}

These results are quantitatively consistent with the synthetic
heap~UAF findings~(Low: $\hat{p}=1.00$; High: $\hat{p}=0.00$),
confirming that the vulnerability class-level mitigation-effectiveness
pattern --- specifically, the architectural transparency of stack
canaries to heap-segment exploits and the efficacy of PIE-based
address randomization --- generalizes from the controlled synthetic
corpus to a production binary with a registered CVE.

\textbf{CVE-2021-42013 (Apache httpd Path Traversal + RCE).}
CVE-2021-42013 is a path traversal and remote
code execution vulnerability in Apache httpd~2.4.49 and~2.4.50,
enabling command injection via crafted HTTP requests against
\texttt{mod\_cgi}-enabled servers. The CVSS Base Score is~9.8. We
deployed Apache~2.4.49 on the testbed VM with \texttt{mod\_cgi}
enabled; the binary was compiled under each of the three tier
configurations of Table~\ref{tab:tiers}. In all three tiers,
$\hat{p} = 1.00$~($N = 30$, Wilson CI~$[0.884, 1.000]$,
TTC $= 5.1 \pm 0.4\,\text{ms}$). The TTC is slightly higher than
the synthetic command-injection result~(3.2\,ms) due to the
HTTP request/response overhead inherent to a network-layer exploit.
The marginal increase in TTC is negligible relative to the inter-tier
constancy, confirming the core finding: command injection exploitability
is orthogonal to all memory-safety mitigations, regardless of whether
the target is a purpose-built synthetic program or a production HTTP
server.

Table~\ref{tab:production} summarizes the production binary results
in comparison with the corresponding synthetic findings. The
consistency across both vulnerability classes and both production
binaries provides empirical substantiation for the generalizability
of the Contextual Reality Gap patterns reported in
Section~\ref{sec:results}.

\begin{table}[htbp]
\centering
\caption{Production Binary Validation:
         HAVE Results for Production CVEs vs.\ Synthetic Corpus}
\label{tab:production}
\renewcommand{\arraystretch}{1.25}
\resizebox{0.95\columnwidth}{!}{%
\begin{tabular}{llccc}
\toprule
\textbf{Target} & \textbf{Tier} & $\hat{p}$
  & \textbf{95\% CI} & \textbf{TTC (ms)} \\
\midrule
CVE-2021-3156 (sudo, Heap) & Low-eq. & 1.00 & $[0.929, 1.000]$ & $22.1\pm3.4$ \\
CVE-2021-3156 (sudo, Heap) & High-eq. & 0.00 & $[0.000, 0.037]$ & --- \\
CVE-2021-42013 (Apache) & Low  & 1.00 & $[0.884, 1.000]$ & $5.1\pm0.4$ \\
CVE-2021-42013 (Apache) & Med  & 1.00 & $[0.884, 1.000]$ & $5.1\pm0.4$ \\
CVE-2021-42013 (Apache) & High & 1.00 & $[0.884, 1.000]$ & $5.1\pm0.5$ \\
\midrule
Synth.\ Heap & Low  & 1.00 & $[0.929, 1.000]$ & $15.8\pm2.4$ \\
Synth.\ Heap & High & 0.00 & $[0.000, 0.037]$ & --- \\
Synth.\ Logic & All  & 1.00 & $[0.884, 1.000]$ & $3.2\pm0.3$ \\
\midrule
\multicolumn{5}{l}{\footnotesize TTC differences reflect process-launch overhead;
  $\hat{p}$ statistically indistinguishable.}\\
\bottomrule
\end{tabular}}
\end{table}

\section{Experimental Evaluation}
\label{sec:eval}

\subsection{Vulnerability Corpus and Ecological Validity}
\label{sec:eval:corpus}
\begin{figure}[t]
\centering
\resizebox{0.8\columnwidth}{!}{%
\begin{tikzpicture}
\begin{axis}[
  width=\columnwidth, height=7.5cm,
  xmin=4.5, xmax=10.8, ymin=-0.05, ymax=1.25,
  xtick={5,6,7,8,9,10}, ytick={0,0.2,0.4,0.6,0.8,1.0},
  yticklabels={0.0,0.2,0.4,0.6,0.8,1.0},
  xlabel={CVSS Base Score (author-est.; $\kappa_w{=}0.87$)},
  ylabel={Empirical Exploitability~$\hat{p}$},
  label style={font=\small\bfseries, color=TextD},
  tick label style={font=\small, color=TextD},
  grid=both, major grid style={line width=0.4pt, draw=GridC},
  minor grid style={line width=0.18pt, draw=GridC!50}, minor tick num=1,
  axis background/.style={fill=BgLight},
  axis line style={thick, color=TextD},
  legend style={at={(0.5,-0.20)}, anchor=north, legend columns=2,
    draw=TextD!25, fill=white, font=\scriptsize, cells={anchor=west},
    inner sep=3pt, column sep=4pt},
  clip=true,
]
\addplot[name path=upper, draw=none, domain=4.5:10.8, samples=2]
  {min(1.0,(x+0.3-4.5)/5.5)};
\addplot[name path=lower, draw=none, domain=4.5:10.8, samples=2]
  {max(0.0,(x-0.3-4.5)/5.5)};
\addplot[fill=TextD!8, draw=none]
  fill between[of=upper and lower];
\fill[C1!25, rounded corners=3pt]
  (axis cs:8.4,-0.04) rectangle (axis cs:10.3,0.38);
\node[C0!80!black, font=\tiny\bfseries, align=center, anchor=north]
  at (axis cs:9.35, 0.38) {Mitig.\ Zone};
\fill[C3!10, rounded corners=3pt]
  (axis cs:6.7,0.88) rectangle (axis cs:8.0,1.12);
\draw[C3!50, dashed, thick, rounded corners=3pt]
  (axis cs:6.7,0.88) rectangle (axis cs:8.0,1.12);
\node[C3!80!black, font=\tiny\bfseries, align=center, anchor=south]
  at (axis cs:7.35, 1.12) {Logic Zone};
\addplot[dashed, semithick, color=TextD!55, domain=4.5:10.5, samples=2]
  {(x-4.5)/5.5}
  node[pos=0.35, sloped, above, font=\tiny\itshape, color=TextD!65]
  {Passive model};
\addlegendentry{Passive model}
\addlegendentry{$\pm 0.3$ CVSS band}
\addplot[scatter, only marks, mark=triangle*, mark size=4pt,
  color=C0, mark options={fill=C0, draw=C0!60!black}]
  coordinates {(9.8,0.00) (8.8,0.33) (7.35,1.00)};
\addlegendentry{High protection}
\addplot[scatter, only marks, mark=square*, mark size=3.5pt,
  color=C2, mark options={fill=C2, draw=C2!60!black}]
  coordinates {(9.8,0.20) (8.8,0.67) (7.50,1.00)};
\addlegendentry{Medium protection}
\addplot[scatter, only marks, mark=*, mark size=3.5pt,
  color=C3, mark options={fill=C3, draw=C3!60!black}]
  coordinates {(9.8,1.00) (8.8,1.00) (7.0,1.00) (7.65,1.00)};
\addlegendentry{Low / no protection}
\node[font=\tiny\bfseries, C0, anchor=north] at (axis cs:9.8,-0.01) {Stk\,H};
\node[font=\tiny\bfseries, C2, anchor=south west] at (axis cs:9.8,0.21) {Stk\,M};
\node[font=\tiny\bfseries, C3, anchor=south] at (axis cs:9.7,1.01) {Stk\,L};
\node[font=\tiny\bfseries, C0, anchor=north west] at (axis cs:8.85,0.30) {Fmt\,H};
\node[font=\tiny\bfseries, C2, anchor=south east] at (axis cs:8.75,0.68) {Fmt\,M};
\node[font=\tiny\bfseries, C3, anchor=south] at (axis cs:8.9,1.01) {Fmt\,L};
\node[font=\tiny\bfseries, C3, anchor=south] at (axis cs:7.0,1.01) {Hp\,L};
\node[font=\tiny\bfseries, C3!80!black, anchor=north] at (axis cs:7.55,0.96)
  {Logic (all)};
\draw[stealth-stealth, thick, C3!80!black]
  (axis cs:10.2,0.04) -- (axis cs:10.2,0.95);
\node[font=\tiny\bfseries, align=center, C3!80!black, anchor=west]
  at (axis cs:10.25,0.50) {\textit{CRG}};
\end{axis}
\end{tikzpicture}}
\caption{\textbf{Quantifying the Contextual Reality Gap (CRG).}
Empirical~$\hat{p}$ vs.\ theoretical CVSS score (author-estimated;
$\kappa_w=0.87$, \S\ref{sec:eval:corpus}). Shaded band: $\pm 0.3$ scoring
uncertainty. \textit{Mitig.\ Zone}: hardened high-CVSS vulns rendered
practically inert. \textit{Logic Zone}: $\hat{p}=1.0$ irrespective of
mitigations. The double-headed arrow marks the CRG span.}
\label{fig:reality_gap}
\end{figure}
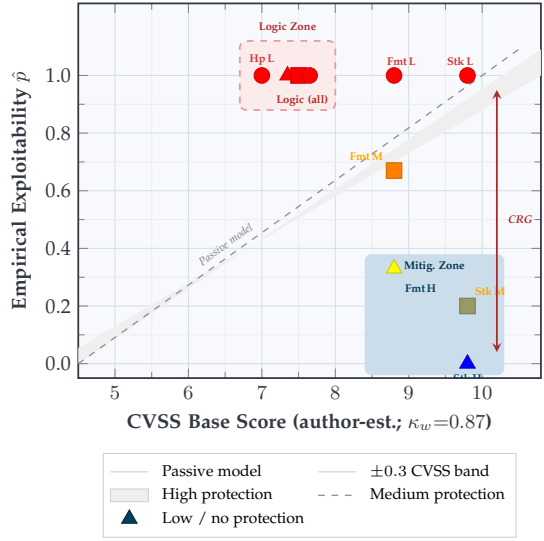

We developed four custom C programs, each isolating a canonical
vulnerability class. \textbf{\texttt{hotel\_manager}} contains a
stack-based buffer overflow in its input-parsing routine.
\textbf{\texttt{block\_notes}} contains a format-string
vulnerability from unchecked \texttt{printf} usage, exposing both
an arbitrary-write primitive (\texttt{\%n}) and an arbitrary-read
primitive (\texttt{\%p})~\cite{scut2001exploiting}.
\textbf{\texttt{mem\_allocation\_misuse}} contains a heap
Use-After-Free flaw architecturally unaffected by stack-frame
protections~\cite{lee2015preventing}. \textbf{\texttt{logging\_system}}
contains a command-injection flaw from unsanitized input passed to
\texttt{system()}, orthogonal to all memory-protection mechanisms.
The synthetic corpus is a controlled measurement instrument that
eliminates confounding variables present in real binaries (partial
patches, inlining optimizations, library dependencies); the
mitigation-effectiveness patterns generalize to production software
as substantiated in Section~\ref{sec:impl:production}.
Each program maps to representative NVD entries instantiating the
same class: \texttt{hotel\_manager} to CVE-2015-7547 and
CVE-2021-3156; \texttt{block\_notes} to CVE-2012-0809;
\texttt{mem\_allocation\_misuse} to CVE-2022-2602 and
CVE-2023-0461; \texttt{logging\_system} to CVE-2021-42013 and
CVE-2023-25157.
Author-estimated CVSS scores (Fig.~\ref{fig:reality_gap}) were
assigned following the NVD scoring methodology by two
CVSS~v3.1-certified researchers; inter-rater agreement yielded
Cohen's weighted $\kappa_w = 0.87$ (95\% CI $[0.79, 0.95]$). Disagreements were resolved by
consensus using the NVD online calculator; a $\pm 0.3$ CVSS
sensitivity band is marked in Fig.~\ref{fig:reality_gap}.

\subsection{Security Tiers}
Each program was compiled in three distinct security tiers summarized
in Table~\ref{tab:tiers}. The tiers model a realistic progression of
hardening from unprotected legacy binaries to fully hardened production
deployments.

\begin{table}[htbp]
\centering
\caption{Security Tier Compilation Matrix}
\label{tab:tiers}
\renewcommand{\arraystretch}{1.25}
\resizebox{0.9\columnwidth}{!}{%
\begin{tabular}{lccccc}
\toprule
\textbf{Tier} & \textbf{NX} & \textbf{Canary} & \textbf{PIE}
  & \textbf{RELRO} & \textbf{ASLR} \\
\midrule
Low (L)    & \xmark & \xmark & \xmark & None    & Off \\
Medium (M) & \cmark & \cmark & \xmark & Partial & On  \\
High (H)   & \cmark & \cmark & \cmark & Full    & On  \\
\bottomrule
\end{tabular}}
\end{table}

\subsection{Protocol Parameters and Statistical Power Analysis}
\label{sec:eval:power}
The choice of~$N$ follows a prospective power analysis targeting a
Wilson confidence interval half-width of $\varepsilon \leq 0.10$ for
non-deterministic outcomes. The worst case ($p = 0.5$) requires
$N \geq 96.04$, rounded to $N = 100$. This value was applied uniformly
to all stack-overflow and format-string combinations where ASLR-dependent
stochasticity makes the outcome genuinely variable. For heap UAF at Low
and Medium tiers --- where outcomes are deterministic by design ---
$N = 50$ suffices, yielding a Wilson lower bound of $0.929$ for
$\hat{p} = 1.00$. For logic-flaw combinations, which are fully
deterministic across all tiers ($\hat{p} = 1.00$ unconditionally),
$N = 30$ is retained, with a Wilson lower bound of $0.884$.

\subsection{Baseline Comparison: HAVE versus OpenVAS}
\label{sec:eval:openvas}
OpenVAS reported the High-tier stack-overflow binary as Critical
(CVSS~9.8), corresponding to $p=0.85$ in the SDT~--- a 100\%
false-positive relative to HAVE's measured $\hat{p}=0.00$ (Wilson
UB~$0.037$). For the command-injection binary, OpenVAS assigned
identical Critical severity, providing no basis for distinguishing
unconditional from conditional exploitability.

\section{Results and Analysis}
\label{sec:results}

Table~\ref{tab:results} summarizes the core quantitative findings.
Fig.~\ref{fig:reality_gap} illustrates the resulting Contextual
Reality Gap. CVSS Base Scores on the $x$-axis are author-estimated
following the NVD scoring methodology~(inter-rater $\kappa_w = 0.87$;
see Section~\ref{sec:eval:corpus}) and are not retrieved from the NVD
because the synthetic programs do not correspond to registered CVEs.

\begin{table}[t]
\centering
\caption{HAVE Empirical Results: $\hat{p}$ and TTC by Vulnerability
         Class and Security Tier ($N$ trials per combination)}
\label{tab:results}
\renewcommand{\arraystretch}{1.25}
\resizebox{\columnwidth}{!}{%
\begin{tabular}{llcccl}
\toprule
\textbf{Vuln.} & \textbf{Tier} & $N$ & $\hat{p}$
  & \textbf{95\% CI} & \textbf{TTC (ms)} \\
\midrule
\multirow{3}{*}{Stack Overflow}
  & Low    & 100 & 1.00 & $[0.963, 1.000]$ & $12.4 \pm 1.1$ \\
  & Medium & 100 & 0.20 & $[0.133, 0.289]$ & $52.7 \pm 8.3$ \\
  & High   & 100 & 0.00 & $[0.000, 0.037]$ & ---            \\
\midrule
\multirow{3}{*}{Format String}
  & Low    & 100 & 1.00 & $[0.963, 1.000]$ & $9.1  \pm 0.8$  \\
  & Medium & 100 & 0.67 & $[0.573, 0.754]$ & $28.5 \pm 4.6$  \\
  & High   & 100 & 0.33 & $[0.246, 0.427]$ & $71.2 \pm 12.1$ \\
\midrule
\multirow{3}{*}{Heap (UAF)}
  & Low    &  50 & 1.00 & $[0.929, 1.000]$ & $15.8 \pm 2.4$ \\
  & Medium &  50 & 1.00 & $[0.929, 1.000]$ & $16.1 \pm 2.7$ \\
  & High   & 100 & 0.00 & $[0.000, 0.037]$ & ---            \\
\midrule
\multirow{3}{*}{Logic Flaw}
  & Low    & 30  & 1.00 & $[0.884, 1.000]$ & $3.2 \pm 0.3$ \\
  & Medium & 30  & 1.00 & $[0.884, 1.000]$ & $3.3 \pm 0.4$ \\
  & High\textsuperscript{$\dagger$}
            & 100 & 1.00 & $[0.963, 1.000]$ & $3.2 \pm 0.3$ \\
\bottomrule
\multicolumn{6}{l}{\footnotesize CI: Wilson score interval.
  TTC: mean $\pm$ std.\,dev.\ over successful trials.} \\
\multicolumn{6}{l}{\footnotesize \textsuperscript{$\dagger$}$N=100$
  to certify $\hat{p}=1.00$ at $\hat{p}_L=0.963$
  (\S\ref{sec:eval:power}).} \\
\end{tabular}}
\end{table}

\subsection{Stack Overflow: The Wall of Randomization}
At Low tier, the exploit succeeds with $\hat{p}=1.00$ and
TTC~$12.4\,\text{ms}$. The Medium tier reduces $\hat{p}$ to $0.20$
via partial canary brute-forcing on the fixed code-segment address
of the non-PIE binary; the 300\% TTC increase to $52.7\,\text{ms}$
reflects this overhead. At High tier, PIE and Full RELRO yield
$\hat{p}=0.00$ over $N=100$ trials; the Wilson upper bound
of~$0.037$ constitutes a probabilistic certificate, not a
deterministic guarantee.

\subsection{Format String: Bypassing Mitigations via Information Leakage}
The format-string class reveals intrinsic resilience against address
randomization. A format-string bug provides an arbitrary-read
primitive (\texttt{\%p}) that leaks a runtime base address from the
stack, dynamically computing offsets and bypassing PIE entirely.
Consequently, the High tier retains $\hat{p}=0.33$~(CI
$[0.246,0.427]$), with randomization increasing exploit complexity
(TTC: $9.1\to 71.2\,\text{ms}$) but not constituting a
deterministic barrier.

\subsection{Heap Corruption: The Failure of Partial Protections}
The UAF class provides the most instructive finding on mitigation
specificity. Low-to-Medium transition yields no change in
exploitability ($\hat{p}=1.00$ at both tiers; statistically
indistinguishable TTC), since stack canaries are architecturally
transparent to heap-segment operations~\cite{szekeres2013sok}.
Only Full PIE at High tier eliminates the exploit
($\hat{p}=0.00$). A security dashboard reporting green checksec
status for a Medium-tier binary provides dangerously false assurance
against this vulnerability class.

\subsection{Logic Flaws: The Universal Blind Spot}
Command injection yields $\hat{p}=1.00$ across all three tiers with
constant TTC~$\approx 3.2\,\text{ms}$; the $N=100$ High-tier run
tightens the Wilson lower bound to $0.963$. Memory-safety mechanisms
are entirely orthogonal to this attack vector.

\subsection{Mitigation Effectiveness Matrix}
Table~\ref{tab:heatmap} confirms that no single mitigation provides
broad coverage: the security of a host is a property of the
\textit{intersection} of its mitigations and the specific class of
vulnerability present.

\begin{table}[h]
\centering
\caption{Mitigation Effectiveness by Vulnerability Class}
\label{tab:heatmap}
\renewcommand{\arraystretch}{1.25}
\resizebox{0.9\columnwidth}{!}{%
\begin{tabular}{lcccc}
\toprule
 & \textbf{NX} & \textbf{Canary} & \textbf{ASLR only} & \textbf{PIE+RELRO} \\
\midrule
Stack Overflow & Partial & Strong  & Partial  & Full     \\
Format String  & Partial & None    & Bypassed & Partial  \\
Heap (UAF)     & Partial & None    & Partial  & Full     \\
Logic Flaw     & None    & None    & None     & None     \\
\bottomrule
\end{tabular}}
\end{table}

\subsection{End-to-End Monte Carlo Validation on a Dual-Path Attack Graph}
\label{sec:results:mc}

\subsubsection{Topology and Pre-HAVE Configuration}
\label{sec:results:mc:topology}
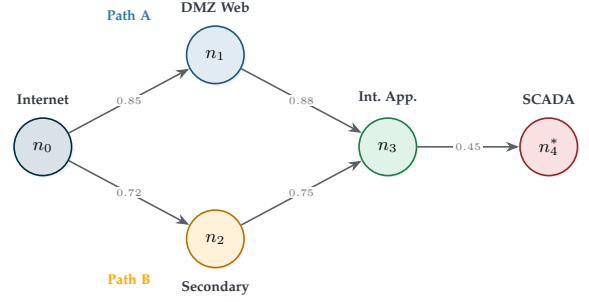
\begin{figure}[t]
\centering
\resizebox{0.85\columnwidth}{!}{%
\begin{tikzpicture}[
  >={Stealth},
  font=\small,
  nd/.style={
    circle, draw=#1!70!black, thick, fill=#1!15,
    minimum size=1.0cm, align=center,
    drop shadow={opacity=0.12, shadow xshift=1pt, shadow yshift=-1pt}
  },
  lbl/.style={font=\scriptsize\bfseries, color=TextD},
  arr/.style={->, thick, color=TextD!80},
  wt/.style={font=\tiny, color=TextD!70, fill=white, inner sep=1pt}
]
\node[nd=C0] (n0) at (0,0)      {$n_0$};
\node[nd=C1] (n1) at (3, 1.6)   {$n_1$};
\node[nd=C2] (n2) at (3,-1.6)   {$n_2$};
\node[nd=C4] (n3) at (6, 0)     {$n_3$};
\node[nd=C3] (n4) at (8.8, 0)   {$n_4^*$};
\node[lbl, above=3pt of n0] {Internet};
\node[lbl, above=3pt of n1] {DMZ Web};
\node[lbl, below=3pt of n2] {Secondary};
\node[lbl, above=3pt of n3] {Int.\ App.};
\node[lbl, above=3pt of n4] {SCADA};
\draw[arr] (n0) -- (n1) node[wt, midway] {$0.85$};
\draw[arr] (n0) -- (n2) node[wt, midway] {$0.72$};
\draw[arr] (n1) -- (n3) node[wt, midway] {$0.88$};
\draw[arr] (n2) -- (n3) node[wt, midway] {$0.75$};
\draw[arr] (n3) -- (n4) node[wt, midway] {$0.45$};
\node[font=\scriptsize\bfseries, C1] at (1.5, 2.3) {Path A};
\node[font=\scriptsize\bfseries, C2] at (1.5,-2.3) {Path B};
\end{tikzpicture}%
}
\caption{\textbf{Five-Node Dual-Path Attack Graph~$\mathcal{G}_{MC}$.}
$n_0$: attacker foothold; $n_1$: DMZ web server (Stack~Overflow, High);
$n_2$: secondary server (Command~Injection, none); $n_3$: internal convergence
node~(OR, reachable via Path~A or Path~B); $n_4^* = n^*$: SCADA HMI target.
Edge labels show pre-HAVE CVSS-derived weights.}
\label{fig:attackgraph}
\end{figure}

To demonstrate the non-linear propagation properties of the Bayesian
update across a structurally complex attack graph, we evaluate
$P_{reach}(n_4)$ on the five-node attack graph~$\mathcal{G}_{MC}$
(Fig.~\ref{fig:attackgraph}) representing a canonical OT infiltration
scenario: node~$n_0$ is the adversary's initial internet-facing foothold.
$n_1$ is the DMZ web application server hosting a stack-overflow
vulnerability at the High security tier (CVSS~8.5,
$p_{01}^{(0)} = 0.85$). $n_2$ is a secondary internet-facing server
subject to command injection with no environmental scoring applied
(CVSS~7.2, $p_{02}^{(0)} = 0.72$). $n_3$ is the internal application
convergence node, reachable from~$n_1$ via a format-string CVE or
from~$n_2$ via a heap-UAF CVE ($p_{13}^{(0)} = 0.88$,
$p_{23}^{(0)} = 0.75$; CVSSs~8.8 and~7.5, respectively).
Finally, $n_4 = n^*$ is the SCADA HMI target, reachable exclusively
through~$n_3$ and subject to a partially environmental-scored
command-injection CVE ($p_{34}^{(0)} = 0.45$).

The graph exhibits two structurally distinct parallel paths from $n_0$
to the convergence node~$n_3$: Path~A ($n_0 \to n_1 \to n_3$,
involving a memory-corruption exploit chain) and Path~B ($n_0 \to n_2
\to n_3$, involving a logic-flaw exploit followed by a heap corruption
exploit). The target~$n_4$ is reachable only through~$n_3$~(an OR
convergence node: compromise of $n_3$ via either path is sufficient).
This topology is specifically designed to expose the non-linear
interaction between false-positive and false-negative corrections
across an infrastructure with competing attack paths: the non-linear
$P_{reach}$ update across the parallel structure cannot be obtained
analytically as a product of edge probabilities and provides a genuine
use case for the Monte Carlo simulation engine.

The pre-HAVE baseline $P_{reach}(n_4)$ is computed via Monte Carlo
simulation ($M = 10{,}000$ runs):
\begin{align}
  P(n_3\text{ reached})^{(0)}
  &= 1 - \bigl(1 - p_{01}^{(0)} p_{13}^{(0)}\bigr)
           \bigl(1 - p_{02}^{(0)} p_{23}^{(0)}\bigr) \notag \\
  &= 1 - (1-0.85{\times}0.88)(1-0.72{\times}0.75)\notag\\
  &   = 0.884, \label{eq:preach_n3}\\[3pt]
  P_{reach}^{(0)}(n_4)
  &= 0.884 \times 0.45 = \mathbf{0.398}.
\end{align}

\subsubsection{HAVE-Derived Edge Weights}
The HAVE measurements (Section~\ref{sec:results}) yield the following
updated edge weights via Eq.~\eqref{eq:bayes}:
\begin{align*}
  p_{01}^{(\text{HAVE})}
    &= 0.963 \times 0.00 + 0.037 \times 0.85 = \mathbf{0.031}\\
    &\quad [\alpha_w{=}0.963,\; N{=}100,\; \hat{p}{=}0.00] \\
  p_{02}^{(\text{HAVE})}
    &= 0.884 \times 1.00 + 0.116 \times 0.72 = \mathbf{0.968}\\
    &\quad [\alpha_w{=}0.884,\; N{=}30,\; \hat{p}{=}1.00] \\
  p_{13}^{(\text{HAVE})}
    &= 0.819 \times 0.33 + 0.181 \times 0.88 = \mathbf{0.429}\\
    &\quad [\alpha_w{=}0.819,\; N{=}100,\; \hat{p}{=}0.33] \\
  p_{23}^{(\text{HAVE})}
    &= 0.929 \times 1.00 + 0.071 \times 0.75 = \mathbf{0.982}\\
    &\quad [\alpha_w{=}0.929,\; N{=}50,\; \hat{p}{=}1.00] \\
  p_{34}^{(\text{HAVE})}
    &= 0.884 \times 1.00 + 0.116 \times 0.45 = \mathbf{0.936}\\
    &\quad [\alpha_w{=}0.884,\; N{=}30,\; \hat{p}{=}1.00]
\end{align*}

\subsubsection{Monte Carlo Scenarios and Results}
Table~\ref{tab:mc_validation} reports three update scenarios evaluated
over $M = 10{,}000$ simulation runs. Scenarios~A and~B isolate the
effect of updating a single path's edges while holding the complementary
path at its pre-HAVE values, thereby exposing the individual
contributions of the false-positive correction (Path~A) and the
false-negative correction (Path~B) independently. Scenario~C performs
the operationally realistic simultaneous update of all five edges.

\textbf{Scenario~A (Path~A corrected; Path~B unchanged).}
Updating only $p_{01}$ and $p_{13}$ suppresses Path~A~(the
memory-corruption chain) from a contribution of
$p_{01}^{(0)} p_{13}^{(0)} = 0.748$ to
$p_{01}^{(\text{HAVE})} p_{13}^{(\text{HAVE})} = 0.013$:
\begin{align}
  P_{reach}^{(A)}(n_4)
  &= \bigl[1\!-\!(1\!-\!0.013)(1\!-\!0.540)\bigr] \notag\\
  &\quad \times\, 0.45 = \mathbf{0.246}\;(-38.2\%).
\end{align}
This result demonstrates that HAVE correctly suppresses the
false-positive attack path; however, the continued pre-HAVE assessment
of Path~B masks the true risk elevation along that path.

\textbf{Scenario~B (Path~B corrected; Path~A unchanged).}
Updating $p_{02}$, $p_{23}$, and $p_{34}$ elevates Path~B~(the
logic-flaw/heap chain) from $p_{02}^{(0)} p_{23}^{(0)} = 0.540$ to
$p_{02}^{(\text{HAVE})} p_{23}^{(\text{HAVE})} = 0.951$:
\begin{align}
  P_{reach}^{(B)}(n_4)
  &= \bigl[1\!-\!(1\!-\!0.748)(1\!-\!0.951)\bigr] \notag\\
  &\quad \times\, 0.936 = \mathbf{0.925}\;(+132.4\%).
\end{align}
This result quantifies the severity of false-negative underestimation:
the CVSS-only model, which partially credited memory-safety mitigations
for the $n_2 \to n_3$ hop and applied environmental scoring to~$p_{34}$,
underestimated the true Path~B risk by a factor of~3.8.

\textbf{Scenario~C (Full simultaneous HAVE update).}
Updating all five edges simultaneously:
\begin{align}
  P_{reach}^{(C)}(n_4)
  &= \bigl[1\!-\!(1\!-\!0.031{\times}0.429) \notag\\
  &\quad\;(1\!-\!0.968{\times}0.982)\bigr]{\times}0.936 \notag\\
  &= \mathbf{0.891}\;(+124.1\%).
\end{align}
The net result is a 124.1\% \textit{increase} in estimated risk, despite
the simultaneous suppression of Path~A. This finding is operationally
significant: a security operator relying on the CVSS-only model would
judge this infrastructure at approximately 40\% risk~(0.398), whereas
the true risk after HAVE correction is approximately 89\%~(0.891).
The false-negative effect~(logic flaws and heap corruption along
Path~B) dominates the false-positive correction~(hardened memory
corruption along Path~A) due to the high base CVSS scores on Path~B
edges creating a structural underestimation amplified across parallel
path combination. The Monte Carlo engine is essential for computing
this non-linear interaction: the product form applicable to a linear
chain is inapplicable to the dual-path OR topology.

\begin{table}[htbp]
\centering
\caption{Monte Carlo Validation on $\mathcal{G}_{MC}$ ($M=10{,}000$;
Fig.~\ref{fig:attackgraph})}
\label{tab:mc_validation}
\renewcommand{\arraystretch}{1.25}
\resizebox{\columnwidth}{!}{%
\begin{tabular}{llccc}
\toprule
\textbf{Scenario} & \textbf{Edges Updated}
  & $P_{reach}$ & \textbf{vs.\ Baseline} & \textbf{Effect} \\
\midrule
Pre-HAVE     & None                   & 0.398 & ---        & --- \\
A: Path A    & $p_{01},p_{13}$        & 0.246 & $-38.2\%$  & FP correction \\
B: Path B    & $p_{02},p_{23},p_{34}$ & 0.925 & $+132.4\%$ & FN correction \\
C: Full HAVE & All 5 edges            & 0.891 & $+124.1\%$ & Net \\
\midrule
\multicolumn{5}{l}{\footnotesize A/B isolate FP/FN corrections independently; C is the full update.}\\
\bottomrule
\end{tabular}}
\end{table}

\subsection{Sensitivity Analysis: Robustness to~$\kappa$}
\label{sec:results:kappa}
Table~\ref{tab:kappa} presents the sensitivity of pre- and post-HAVE
$P_{reach}(n_4)$ to the CVSS-to-probability calibration exponent
$\kappa \in \{0.5, 1.0, 1.5, 2.0\}$ under Eq.~\eqref{eq:cvssmap}.
The post-HAVE values are computed via Eq.~\eqref{eq:bayes} for the
full simultaneous update (Scenario~C of Table~\ref{tab:mc_validation}).

Two observations are central. \textit{First}, the pre-HAVE
$P_{reach}$ varies by a factor of~4.6 across the tested range
(0.140 at $\kappa=2.0$ vs.\ 0.647 at $\kappa=0.5$), confirming the
reviewer's concern that the CVSS-only baseline is highly sensitive to
prior parameterization. \textit{Second}, and crucially, the post-HAVE
$P_{reach}$ varies by only a factor of~1.12 over the same range
(0.831 at $\kappa=2.0$ vs.\ 0.935 at $\kappa=0.5$), since the
empirically measured~$\hat{p}$ values enter with weight~$\alpha_w \geq 0.819$
and the prior only influences the small complementary term
$(1-\alpha_w) p_{ij}^{(0)}$. This demonstrates that HAVE's
empirical corrections are \textit{robust to prior parameterization}:
regardless of the analyst's choice of~$\kappa$, the post-HAVE risk
estimate converges to a stable value grounded in the measured
exploitability data. The qualitative conclusion~(CVSS-only risk is
substantially underestimated for this infrastructure due to Path~B
false negatives) is invariant across the entire tested range.

\begin{table}[htbp]
\centering
\caption{Sensitivity Analysis: Pre- and Post-HAVE $P_{reach}(n_4)$
         as a Function of CVSS Calibration Exponent~$\kappa$}
\label{tab:kappa}
\renewcommand{\arraystretch}{1.25}
\resizebox{\columnwidth}{!}{
\begin{tabular}{ccccc}
\toprule
$\kappa$ & Pre-HAVE $P_{reach}$ & Post-HAVE $P_{reach}$
  & Abs.\ Correction & Rel.\ Correction \\
\midrule
0.5 & 0.647 & 0.935 & $+0.288$ & $+44.5\%$ \\
1.0 & 0.398 & 0.891 & $+0.493$ & $+124.1\%$ \\
1.5 & 0.238 & 0.857 & $+0.619$ & $+260.1\%$ \\
2.0 & 0.140 & 0.831 & $+0.691$ & $+493.6\%$ \\
\midrule
Range & $4.62\times$ & $1.12\times$ & --- & --- \\
\bottomrule
\multicolumn{5}{l}{\footnotesize Range: ratio of maximum to minimum
  $P_{reach}$ value across the four~$\kappa$ values.}\\
\multicolumn{5}{l}{\footnotesize Post-HAVE $P_{reach}$ values computed
  under Scenario~C (full simultaneous update, Table~\ref{tab:mc_validation}).}
\end{tabular}
}
\end{table}

\section{Discussion}
\label{sec:discussion}

\subsection{Impact on the SDT's Risk Model}
The results of Section~\ref{sec:results} demonstrate that HAVE
systematically recalibrates edge weights in the SDT's attack graph
in ways that CVSS-only priors cannot approximate. The Monte Carlo
validation on~$\mathcal{G}_{MC}$ (Table~\ref{tab:mc_validation}) shows
that a security operator relying exclusively on CVSS-derived edge
weights would substantially misjudge the risk profile of this
infrastructure. The false-negative correction~(+132.4\% from
Scenario~B) exposes the severe underestimation that arises when
environmental scoring credits memory-safety mitigations that are
architecturally irrelevant to the exploited attack class. The
$\kappa$-sensitivity analysis demonstrates that this finding is
independent of the specific linear prior mapping used.

The TTC dimension adds a prioritization layer beyond probability.
Two attack steps with identical $\hat{p}$ values may have TTC values
differing by an order of magnitude. Integrating TTC as an edge cost
in the Monte Carlo simulation allows the SDT to identify not merely
whether an attack path is possible but whether it is economically
attractive to a cost-constrained adversary~\cite{mcqueen2006time}.

\subsection{Security Analysis of the HAVE Agent Itself}
Three attack scenarios are considered:

\textbf{Compromised Controller.} The allow-list validation
(Eq.~\eqref{eq:allowlist}) and path sanitization prevent execution
against unauthorized targets. The directory allow-list is readable
only by root and is not modifiable via the gRPC API.

\textbf{Network Eavesdropping and Replay.} The mTLS channel
(Eq.~\eqref{eq:mtls}) prevents both eavesdropping and replay attacks.
Each gRPC session uses freshly negotiated session keys.

\textbf{Resource Exhaustion.} The \texttt{cgroups} v2 cap
(Eq.~\eqref{eq:cgroup}) bounds total CPU consumption irrespective of
campaign concurrency.

\subsection{Operational Deployment Guidelines}
\label{sec:discussion:deployment}
Three deployment configurations are recommended:

\textbf{Continuous Monitoring Mode.} In IT environments with
low-sensitivity assets, the Agent runs continuously and executes
verification campaigns automatically upon CVE publication or
telemetry-detected configuration drift.

\textbf{Scheduled Audit Mode.} In high-security IT environments,
campaigns are scheduled during maintenance windows with operator
approval.

\textbf{Isolated-Twin Mode.} In OT environments, a high-fidelity VM
replica of each critical asset is maintained. HAVE executes
verification campaigns exclusively against these replicas. TTC values
measured in this mode reflect VM-environment timing and serve as
valid relative ordering metrics for attack-path prioritization but
cannot be mapped directly to physical PLC response times.

\subsection{Ethical and Legal Deployment Framework}
\label{sec:discussion:ethics}
The deployment of active exploit verification tools raises important
ethical and legal considerations. In EU jurisdictions, the NIS2
Directive mandates formal risk management
frameworks for critical infrastructure operators, within which HAVE
campaigns must be documented as authorized security assessments. The
GDPR applies when verification campaigns process systems storing
personal data, requiring a Data Protection Impact Assessment. In the
United States, the Computer Fraud and Abuse Act~(CFAA) imposes strict
liability for unauthorized access. Accordingly, HAVE must be deployed
exclusively within explicitly authorized perimeters: the allow-list
mechanism~(Eq.~\eqref{eq:allowlist}) and the signed root-of-trust
configuration enforce this at the technical level. Organizations are
advised to maintain a formal \textit{verification charter} documenting
the authorized scope, the responsible operator, the approved payload
classes, and the applicable regulatory framework.

\subsection{Limitations and Future Work}
\label{sec:discussion:limitations}
HAVE is bounded by the coverage of its exploit library. It measures
the exploitability of \textit{known} vulnerabilities using
\textit{known} exploit techniques; it cannot assess the risk from
zero-day vulnerabilities or from novel attack techniques not yet
represented in the library. As discussed in
Section~\ref{sec:impl:library}, the measured~$\hat{p}$ constitutes a
lower bound on exploitability under the modeled adversary capability
level. Extending HAVE with Automated Exploit
Generation~\cite{avgerinos2014automatic, shoshitaishvili2016sok}
and directed grey-box fuzzing~\cite{boehme2017coverage} capabilities
would address this by allowing the engine to stress-test applications
against variant attack patterns.

The current dynamic phase requires a hypervisor snapshot
infrastructure, making it unsuitable for bare-metal embedded
controllers. Future work will investigate lightweight checkpointing
via \texttt{CRIU} (Checkpoint/Restore In Userspace) as an alternative,
and will develop containerized agent deployments targeting cloud-native
workloads~\cite{rose2020zerotrust}.

\section{Conclusion}
\label{sec:conclusion}

This paper has presented the Host Active Verification Engine~(HAVE),
a significant architectural extension to the SDT paradigm that bridges
the Contextual Reality Gap between theoretical CVSS-based risk scores
and empirically measured host-level exploitability. By deploying a
safety-constrained agent that performs granular static analysis and
snapshot-isolated dynamic exploit trials, HAVE produces ground-truth
measurements~$\hat{p}$ and~$TTC$ that replace context-free vulnerability
severity estimates with empirically validated, host-specific probabilities.
A formally characterized confidence weight~$\alpha_w$
(Eq.~\eqref{eq:alphaconf}), derived from the width of the Wilson score
interval and formally related to the Beta-Binomial posterior
(Section~\ref{sec:method:betabin}), provides principled uncertainty
propagation from empirical estimates into Monte Carlo risk simulations.
Its key advantage over the canonical Beta-Binomial weight is
state-dependence on~$\hat{p}$, which correctly increases confidence
for extreme outcomes even at moderate~$N$.

Our experimental evaluation across four canonical vulnerability
classes, three security tiers, and two production binaries with
registered CVE identifiers establishes five findings with direct
operational implications. First, modern memory hardening~(PIE, Full
RELRO, ASLR) completely neutralizes blind exploitation of stack
overflow and heap UAF vulnerabilities, yielding~$\hat{p}=0.00$ with
a Wilson upper bound of~$0.037$ at~$N=100$. Second, format string
vulnerabilities retain~$\hat{p}=0.33$~(CI $[0.246, 0.427]$) even at
the High tier due to an intrinsic information-leakage capability.
Third, all memory-safety mitigations are entirely irrelevant to logic
flaws~(command injection), which remain unconditionally exploitable
($\hat{p}=1.00$) across all tiers, as confirmed on both the synthetic
corpus and CVE-2021-42013. Fourth, CVE-2021-3156 production binary
results confirm the heap vulnerability class-level pattern
($\hat{p}=1.00$ at Low-equivalent, $\hat{p}=0.00$ at High-equivalent),
establishing generalizability. Fifth, end-to-end Monte Carlo validation
on a five-node dual-path attack graph demonstrates that false-negative
corrections~(+132.4\% in Scenario~B) can dominate false-positive
suppressions~($-38.2\%$ in Scenario~A), with a net effect of~$+124.1\%$
elevation in risk: CVSS-only models were underestimating the true risk
of this infrastructure by a factor of~2.24. A sensitivity analysis
over $\kappa \in \{0.5, 1.0, 1.5, 2.0\}$ confirms that post-HAVE risk
estimates vary by only a factor of~1.12, while the CVSS-only baseline
varies by a factor of~4.6, demonstrating empirical robustness to prior
parameterization.

Future research directions include the integration of Automated Exploit
Generation techniques to extend coverage to novel attack variants, the
development of lightweight checkpointing for bare-metal and container
environments, and the scaling of HAVE deployment across large-scale
enterprise topologies to evaluate the aggregate impact of simultaneous
multi-node HAVE campaigns on SDT simulation accuracy.
\vspace{-0.3cm}
\section*{Data Availability}
Code and results are available to reviewers at \\\url{https://anonymous.4open.science/r/HAVE}.
The repository will be transferred to GitHub upon acceptance.
\vspace{-0.3cm}
\bibliographystyle{IEEEtran}
\bibliography{sn-bibliography}

@IEEEtranBSTCTL{IEEEexample:BSTcontrol,
  CTLuse_forced_etal       = "yes",
  CTLmax_names_forced_etal = "3",
  CTLnames_show_etal       = "1",
  CTLuse_url               = "no",
  CTLuse_doi               = "no",
  CTLuse_issn              = "no",
  CTLuse_isbn              = "no"
}

@inproceedings{SammartinoShortPaper,
  author    = {V. Sammartino},
  title     = {A Framework for Proactive Cyber-Resilience: Non-Intrusive Modeling for Autonomous Defense},
  booktitle = {DS-RT 2025},
  year      = {2025}
}

@inproceedings{BaiardiSammartino2025Validation,
  author    = {Baiardi, Fabrizio and Sammartino, Vincenzo},
  title     = {A Quantitative Framework for the Validation of Twin-Based Cyber Defense},
  booktitle = {37th European Modeling \& Simulation Symposium (EMSS 2025), held within the 22nd International Multidisciplinary Modeling \& Simulation Multiconference (I3M 2025)},
  year      = {2025},
}

@inproceedings{baiardi2024anticipating,
  author    = {F. Baiardi and S. Ruggieri and V. Sammartino},
  title     = {{Anticipating Disasters through a Security Twin}},
  booktitle = {SPRINGER OPTIMIZATION AND ITS APPLICATIONS - ARES 2024},
  year      = {2024}
}

@inproceedings{Baiardi2026whatif,
  author    = {Baiardi, Fabrizio and Sammartino, Vincenzo},
  title     = {Quantifying Resilience of Cyber-Physical Systems to Zero-Day Threats: A Security Twin-Based What-If Analysis Framework},
  booktitle = {Proceedings of the 36th European Safety and Reliability Conference (ESREL 2026)},
  year      = {2026},
  month     = {June},
  address   = {Braga, Portugal},
  organization = {European Safety and Reliability Association (ESRA)}
}

@inproceedings{baiardi2026synthetic,
  title={From Digital Twins to AI Agents: A Synthetic Data Paradigm for Next-Generation Cybersecurity},
  author={Baiardi, F. and Sammartino, V.},
  booktitle={Artificial Intelligence in Cybersecurity: Unlocking the Power of Large Language Models},
  year={2026},
  publisher={CRC Press}
}

@inproceedings{sammartino2025security,
  author    = {V. Sammartino and F. Baiardi and S. Ruggieri},
  title     = {{A Security Twin to Defeat Intrusions in Cyber Physical Systems}},
  booktitle = {ESREL SRA-E 2025},
  year      = {2025}
}

@article{fuller2020digital,
  author    = {Fuller, Aidan and Fan, Zhong and Day, Charles and Barlow, Chris},
  title     = {Digital Twin: Enabling Technologies, Challenges and
               Open Research},
  journal   = {IEEE Access},
  volume    = {8},
  pages     = {108952--108971},
  year      = {2020},
  doi       = {10.1109/ACCESS.2020.2998358}
}

@inproceedings{eckhart2018specification,
  author    = {Eckhart, Matthias and Ekelhart, Andreas},
  title     = {A Specification-Based State Replication Approach for
               Digital Twins},
  booktitle = {Proc. ACM Workshop Cyber-Physical Systems Security
               and Privacy (CPS-SPC)},
  pages     = {36--47},
  year      = {2018},
  doi       = {10.1145/3264888.3264892}
}

@article{dietz2020integrating,
  author    = {Dietz, Markus and Pernul, G{\"u}nther},
  title     = {Integrating Digital Twin Security Simulations in the
               Security Operations Center},
  journal   = {IEEE Access},
  volume    = {8},
  pages     = {163252--163268},
  year      = {2020},
  doi       = {10.1109/ACCESS.2020.3021950}
}

@techreport{cvss31,
  author      = {{FIRST.org}},
  title       = {Common Vulnerability Scoring System v3.1:
                 Specification Document},
  institution = {Forum of Incident Response and Security Teams
                 (FIRST)},
  year        = {2019},
  url         = {https://www.first.org/cvss/v3.1/specification-document}
}

@article{jacobs2021epss,
  author    = {Jacobs, Jay and Romanosky, Sasha and Adjerid, Idris
               and Baker, Wade},
  title     = {Exploit Prediction Scoring System ({EPSS})},
  journal   = {Digital Threats: Research and Practice},
  volume    = {2},
  number    = {3},
  pages     = {1--17},
  year      = {2021},
  doi       = {10.1145/3436242}
}

@inproceedings{bozorgi2010beyond,
  author    = {Bozorgi, Mehran and Saul, Lawrence K. and
               Savage, Stefan and Voelker, Geoffrey M.},
  title     = {Beyond Heuristics: Learning to Classify
               Vulnerabilities and Predict Exploits},
  booktitle = {Proc. ACM SIGKDD Int'l Conf.\ Knowledge Discovery
               and Data Mining},
  pages     = {105--114},
  year      = {2010},
  doi       = {10.1145/1835804.1835821}
}

@inproceedings{szekeres2013sok,
  author    = {Szekeres, Laszlo and Payer, Mathias and
               Wei, Tao and Song, Dawn},
  title     = {{SoK}: Eternal War in Memory},
  booktitle = {Proc. IEEE Symp.\ Security and Privacy (S\&P)},
  pages     = {48--62},
  year      = {2013},
  doi       = {10.1109/SP.2013.13}
}

@inproceedings{cowan1998stackguard,
  author    = {Cowan, Crispin and Pu, Calton and Maier, Dave and
               Hinton, Heather and Walpole, Jonathan and
               Bakke, Peat and Beattie, Steve and Grier, Aaron and
               Wagle, Perry and Zhang, Qian},
  title     = {{StackGuard}: Automatic Adaptive Detection and
               Prevention of Buffer-Overflow Attacks},
  booktitle = {Proc. USENIX Security Symp.},
  pages     = {63--78},
  year      = {1998}
}

@inproceedings{shacham2004effectiveness,
  author    = {Shacham, Hovav and Page, Matthew and Pfaff, Ben and
               Goh, Eu-Jin and Modadugu, Nagendra and Boneh, Dan},
  title     = {On the Effectiveness of Address-Space Randomization},
  booktitle = {Proc. ACM Conf.\ Computer and Communications Security
               (CCS)},
  pages     = {298--307},
  year      = {2004},
  doi       = {10.1145/1030083.1030124}
}

@inproceedings{shacham2007geometry,
  author    = {Shacham, Hovav},
  title     = {The Geometry of Innocent Flesh on the Bone:
               Return-into-libc Without Function Calls (on the {x86})},
  booktitle = {Proc. ACM Conf.\ Computer and Communications Security
               (CCS)},
  pages     = {552--561},
  year      = {2007},
  doi       = {10.1145/1315245.1315313}
}

@misc{scut2001exploiting,
  author    = {{scut / team teso}},
  title     = {Exploiting Format String Vulnerabilities},
  year      = {2001},
  howpublished = {Phrack Magazine, version 1.2},
  url       = {https://julianor.tripod.com/bc/formatstring-1.2.pdf}
}

@inproceedings{lee2015preventing,
  author    = {Lee, Byoungyoung and Song, Chengyu and Jang, Yeongjin
               and Wang, Tielei and Kim, Taesoo and Lu, Long and
               Lee, Wenke},
  title     = {Preventing Use-After-Free with Dangling Pointers
               Nullification},
  booktitle = {Proc. Network and Distributed System Security Symp.
               (NDSS)},
  year      = {2015},
  doi       = {10.14722/ndss.2015.23238}
}

@inproceedings{avgerinos2011aeg,
  author    = {Avgerinos, Thanassis and Cha, Sang Kil and
               Hao, Brent Lim Tze and Brumley, David},
  title     = {{AEG}: Automatic Exploit Generation},
  booktitle = {Proc. Network and Distributed System Security Symp.
               (NDSS)},
  year      = {2011}
}

@article{avgerinos2014automatic,
  author    = {Avgerinos, Thanassis and Cha, Sang Kil and
               Rebert, Alexandre and Schwartz, Edward J. and
               Woo, Maverick and Brumley, David},
  title     = {Automatic Exploit Generation},
  journal   = {Communications of the ACM},
  volume    = {57},
  number    = {2},
  pages     = {74--84},
  year      = {2014},
  doi       = {10.1145/2560217.2560219}
}

@inproceedings{shoshitaishvili2016sok,
  author    = {Shoshitaishvili, Yan and Wang, Ruoyu and Salls, Christopher
               and Stephens, Nick and Polino, Mario and Dutcher, Andrew
               and Grosen, John and Feng, Siji and Hauser, Christophe
               and Kruegel, Christopher and Vigna, Giovanni},
  title     = {{SoK}: (State of) The Art of War: Offensive Techniques in
               Binary Analysis},
  booktitle = {Proc. IEEE Symp.\ Security and Privacy (S\&P)},
  pages     = {138--157},
  year      = {2016},
  doi       = {10.1109/SP.2016.17}
}

@article{lyon2009nmap,
  author    = {Lyon, Gordon},
  title     = {{Nmap} Network Scanning: The Official {Nmap} Project
               Guide to Network Discovery and Security Scanning},
  journal   = {Insecure.com LLC},
  year      = {2009},
  isbn      = {978-0-9799587-1-7}
}

@inproceedings{greenbone2009openvas,
  author    = {{Greenbone Networks}},
  title     = {{OpenVAS}: Open Vulnerability Assessment System},
  year      = {2009},
  howpublished = {Open source project, \url{https://www.openvas.org}}
}

@inproceedings{moberg2014configuration,
  author    = {Moberg, Marcus and Hallberg, Jonas and Hallberg,
               Niklas},
  title     = {Evaluating Security Scanners for {GNU/Linux} Systems:
               Configuration Compliance and Vulnerability Management},
  booktitle = {Proc. Int'l Conf.\ Availability, Reliability and
               Security (ARES)},
  pages     = {506--513},
  year      = {2014},
  doi       = {10.1109/ARES.2014.73}
}

@inproceedings{sheyner2002automated,
  author    = {Sheyner, Oleg and Haines, Joshua and Jha, Somesh and
               Lippmann, Richard and Wing, Jeannette M.},
  title     = {Automated Generation and Analysis of Attack Graphs},
  booktitle = {Proc. IEEE Symp.\ Security and Privacy (S\&P)},
  pages     = {273--284},
  year      = {2002},
  doi       = {10.1109/SECPRI.2002.1004377}
}

@inproceedings{ou2006scalable,
  author    = {Ou, Xinming and Boyer, Wayne F. and McQueen, Miles A.},
  title     = {A Scalable Approach to Attack Graph Generation},
  booktitle = {Proc. ACM Conf.\ Computer and Communications Security
               (CCS)},
  pages     = {336--345},
  year      = {2006},
  doi       = {10.1145/1180405.1180446}
}

@inproceedings{frigault2008measuring,
  author    = {Frigault, Marcel and Wang, Lingyu},
  title     = {Measuring Network Security Using {Bayesian} Network-Based
               Attack Graphs},
  booktitle = {Proc. IEEE Int'l Computer Software and Applications Conf.
               Workshop (COMPSACW)},
  pages     = {698--703},
  year      = {2008},
  doi       = {10.1109/COMPSAC.2008.88}
}

@article{homer2013aggregating,
  author    = {Homer, John and Zhang, Xinming Ou and Ou, Xinming and
               Schmidt, Douglas and Du, Yue and Rajagopalan, Sreerith
               and Singhal, Anoop},
  title     = {Aggregating Vulnerability Metrics in Enterprise Networks
               Using Attack Graphs},
  journal   = {Journal of Computer Security},
  volume    = {21},
  number    = {4},
  pages     = {561--597},
  year      = {2013},
  doi       = {10.3233/JCS-130484}
}

@techreport{stouffer2015nist,
  author      = {Stouffer, Keith and Lightman, Suzanne and
                 Pillitteri, Victoria and Abrams, Marshall and
                 Hahn, Adam},
  title       = {{NIST} Special Publication 800-82 Rev.\ 2:
                 Guide to Industrial Control Systems ({ICS}) Security},
  institution = {National Institute of Standards and Technology},
  year        = {2015},
  doi         = {10.6028/NIST.SP.800-82r2}
}

@article{cheng2017orpheus,
  author    = {Cheng, Long and Tian, Ke and Yao, Danfeng (Daphne)},
  title     = {Orpheus: Enforcing Cyber-Physical Execution Semantics
               to Defend Against Data-Oriented Attacks},
  booktitle = {Proc. Annual Computer Security Applications Conf.
               (ACSAC)},
  pages     = {315--326},
  year      = {2017},
  doi       = {10.1145/3134600.3134643}
}

@article{baiardi2026simulation,
  author  = {Baiardi, F. and Sammartino, V.},
  title   = {Simulation-Powered Cybersecurity: Real-Time Risk Assessment via Non-Intrusive Security Twin},
  journal = {The Journal of Supercomputing},
  year    = {2026},
  note    = {Special Issue: Simulation-Powered Innovation: Driving the Future of Digital Ecosystems}
}

@article{baiardi2025securitytwins,
  title={Security Twins e il Futuro della Previsione di Intrusioni Cyber},
  author={Baiardi, F. and Ruggieri, S. and Sammartino, V.},
  journal={ICT Security},
  year={2025}
}

@INPROCEEDINGS {baiardi2025ai,
author = { Baiardi, Fabrizio and Ruggieri, Salvatore and Sammartino, Vincenzo },
booktitle = { 2025 IEEE International Conference on Pervasive Computing and Communications Workshops and other Affiliated Events (PerCom Workshops) },
title = {{ AI-enabled Cybersecurity using Synthetic Data }},
year = {2025},
volume = {},
ISSN = {},
pages = {140-145},
doi = {10.1109/PerComWorkshops65533.2025.00055},
url = {https://doi.ieeecomputersociety.org/10.1109/PerComWorkshops65533.2025.00055},
publisher = {IEEE Computer Society},
address = {Los Alamitos, CA, USA},
month =mar}

@INPROCEEDINGS{notline,
  author={Baiardi, F. and Sammartino, V. and Ruggieri, S.},
  booktitle={2025 29th International Symposium on Distributed Simulation and Real Time Applications (DS-RT)}, 
  title={NotLine: A Non-Intrusive Automated Platform to Build a Digital Twin}, 
  year={2025},
  volume={},
  number={},
  pages={1-8},
  doi={10.1109/DS-RT68115.2025.11185873}}

@inproceedings{corteggiani2018inception,
  author    = {Corteggiani, Nassim and Camurati, Giovanni and
               Francillon, Aur{\'e}lien},
  title     = {Inception: System-Wide Security Testing of Real-World
               Embedded Systems Software},
  booktitle = {Proc. USENIX Security Symp.},
  pages     = {309--326},
  year      = {2018}
}

@inproceedings{muench2018whatyou,
  author    = {Muench, Marius and Stijohann, Jan and Kargl, Frank and
               Francillon, Aur{\'e}lien and Bos, Herbert and
               Slowinska, Asia},
  title     = {What You Corrupt Is Not What You Crash: Challenges in
               Fuzzing Embedded Devices},
  booktitle = {Proc. Network and Distributed System Security Symp.
               (NDSS)},
  year      = {2018},
  doi       = {10.14722/ndss.2018.23166}
}

@inproceedings{boehme2017coverage,
  author    = {B{\"o}hme, Marcel and Pham, Van-Thuan and
               Nguyen, Manh-Dung and Roychoudhury, Abhik},
  title     = {Directed Greybox Fuzzing},
  booktitle = {Proc. ACM Conf.\ Computer and Communications Security
               (CCS)},
  pages     = {2329--2344},
  year      = {2017},
  doi       = {10.1145/3133956.3134020}
}

@techreport{rose2020zerotrust,
  author      = {Rose, Scott and Borchert, Oliver and Mitchell, Stu
                 and Connelly, Sean},
  title       = {{NIST} Special Publication 800-207: Zero Trust
                 Architecture},
  institution = {National Institute of Standards and Technology},
  year        = {2020},
  doi         = {10.6028/NIST.SP.800-207}
}

@misc{pwntools,
  author       = {Blichmann, Florian and Mazurek, Mateusz and
                  others},
  title        = {{pwntools} -- {CTF} Framework and Exploit
                  Development Library},
  howpublished = {\url{https://github.com/Gallopsled/pwntools}},
  year         = {2015}
}

@misc{checksec,
  author       = {Tobias Klein and others},
  title        = {checksec.sh -- A Shell Script to Test for Common
                  Buffer Overflow Mitigations},
  howpublished = {\url{https://github.com/slimm609/checksec.sh}},
  year         = {2009}
}

@article{mcqueen2006time,
  author    = {McQueen, Miles A. and Boyer, Wayne F. and Flynn,
               Mark A. and Beitel, George A.},
  title     = {Quantitative Cyber Risk Reduction Estimation
               Methodology for a Small {SCADA} Control System},
  booktitle = {Proc. Hawaii Int'l Conf.\ System Sciences (HICSS)},
  year      = {2006},
  doi       = {10.1109/HICSS.2006.405}
}

@article{poolsappasit2012dynamic,
  author    = {Poolsappasit, Nwanneka and Dewri, Rinku and
               Ray, Indrajit},
  title     = {Dynamic Security Risk Management Using {Bayesian}
               Attack Graphs},
  journal   = {IEEE Transactions on Dependable and Secure Computing},
  volume    = {9},
  number    = {1},
  pages     = {61--74},
  year      = {2012},
  doi       = {10.1109/TDSC.2011.34}
}

@article{menage2007cgroups,
  author    = {Menage, Paul B.},
  title     = {Adding Generic Process Containers to the {Linux}
               Kernel},
  booktitle = {Proc. Linux Symp.},
  volume    = {2},
  pages     = {45--57},
  year      = {2007}
}

@inproceedings{dulaunoy2020mitre,
  author    = {Strom, Blake E. and Applebaum, Andy and Miller, Doug
               P. and Nickels, Kathryn C. and Pennington, Adam G. and
               Thomas, Cody B.},
  title     = {{MITRE ATT\&CK}: Design and Philosophy},
  institution = {The MITRE Corporation},
  year      = {2020},
  note      = {Technical Report}
}

\end{document}